\documentstyle[12pt,aaspp4]{article}
\begin{document}
\parindent=1.0cm

\title{A NEAR-INFRARED PHOTOMETRIC STUDY OF THE LOW LATITUDE GLOBULAR CLUSTERS 
LILLER 1, DJORGOVSKI 1, HP 1, and NGC 6528}

\author{T. J. Davidge \altaffilmark{1} , \altaffilmark{2}}

\affil{Canadian Gemini Office, Herzberg Institute of Astrophysics, 
\\National Research Council of Canada, 5071 W. Saanich Road,\\ Victoria
B.C. Canada V8X 4M6\\ {\it email:tim.davidge@hia.nrc.ca}}

\altaffiltext{1}{Visiting Astronomer, Canada-France-Hawaii Telescope, which is 
operated by the National Research Council of Canada, the Centre National de la 
Recherche Scientifique, and the University of Hawaii}

\altaffiltext{2}{Visiting Astronomer, Cerro Tololo Inter-American Observatory, 
which is operated by AURA Inc. under contract from the National Science 
Foundation}

\begin{abstract}

	Images recorded through $J, H, Ks, 2.2\mu$m continuum, and $2.3\mu$m 
CO filters are used to investigate the stellar contents of the low Galactic 
latitude globular clusters NGC 6528, Liller 1, Djorgovski 1, and HP 1, 
as well as surrounding bulge fields. Metallicities are 
estimated for the latter three clusters by comparing the colors 
and CO indices of giant branch stars with those in other clusters and the 
bulge, while reddenings are estimated from the colors of 
bright bulge stars in the surrounding fields. In some cases the metallicities 
and reddenings are significantly different from previous estimates. 

	The horizontal branch (HB) in NGC 6528 occurs at $K = 13.2$, and 
the $K$ luminosity function of HB stars in this cluster is not significantly 
different from those in the moderately metal-rich clusters NGC 6304 and 
NGC 6316 (Davidge et al. 1992, ApJS, 81, 251). R', the ratio of HB to bright 
red giant branch (RGB) and asymptotic giant branch (AGB) stars, is $1.25 \pm 
0.25$, in good agreement with other globular clusters. The RGB 
bump in NGC 6528 is detected at $K = 13.8$, and the relative 
brightness of this feature with respect to the HB is consistent with 
the cluster being very metal-rich.

	The HB in Liller 1 occurs at $K = 14.4$. Stars in Liller 1 have smaller 
$J-K$ colors and CO indices than bulge giants with the same brightness, and a 
comparison between the Liller 1 and NGC 6528 sequences on the 
$(K, J-K)$ and $(K, CO)$ color-magnitude diagrams (CMDs) 
indicates that Liller 1 has a metallicity comparable to, and perhaps even 
slightly lower than, that of NGC 6528. Stars in Liller 1 and NGC 6528 occupy 
the same region of the $(J-H, H-K)$ two-color diagram as bulge giants, which is 
consistent with these clusters forming as part of the bulge, as suggested by 
Minniti (1995a, AJ, 109, 1663).

	Stars in Djorgovski 1 and HP 1 have significantly smaller CO indices 
than stars with the same brightness in the surrounding bulge fields, and CMDs 
for these clusters are constructed by selecting objects according to CO index. 
After adjusting for differences in distance and reddening, the Djorgovski 1 
giant branch on the $(K, J-K)$ CMD is well-matched by that of the [Fe/H] $= 
-2.2$ cluster M92, while the giant branch of HP 1 follows that of the 
[Fe/H] $= -1.6$ globular cluster M13. The metallicities infered for 
both clusters are consistent with integrated $J-K$ colors and CO indices 
measured through moderately large ($\sim 35 - 45$ arcsec) apertures. 
Based on the brightness of the RGB-tip it is concluded that Djorgovski 1 has 
a distance modulus $\mu_0 = 15.4$, while the distance modulus of HP 1 is 
$\mu_0 = 14.7$. 

	The brightest stars in the fields surrounding 
Liller 1, Djorgovski 1, and HP 1 follow the same 
relation between M$_K$ and CO index as M giants in Baade's Window 
(BW). Stars as faint as the HB have been detected in the field surrounding 
HP 1, and the relative numbers of HB and upper giant branch stars indicate that 
R' $= 1.14 \pm 0.11$. This relatively low value reinforces earlier studies 
which found that R', and hence Helium abundance, likely does not vary with 
radius in the innermost regions of the Galactic bulge.

\vspace{0.3cm}

\end{abstract}

\section{INTRODUCTION}

	The spatial distribution and kinematic 
properties of globular clusters show a metallicity dependance (e.g. 
Kinman 1959, Zinn 1985) that was imprinted during the early evolution 
of the Galaxy. Globular clusters with [Fe/H] $\geq -1$ 
tend to be seen at low Galactic latitudes, and it has been 
suggested that these objects may have formed as part of the thick disk (Zinn 
1985, Armandroff 1989), the bulge (Minitti 1995a), and/or 
the bar (Burkert \& Smith 1997). Consequently, studies of these 
objects have the potential to provide insight into the early evolution of 
the inner Galaxy. On the other hand, clusters with 
[Fe/H] $\leq -1$ occur in diverse environments spanning a range of 
galactocentric distances, and the relationship between metal-poor clusters in 
the outer halo and inner spheroid is not clear. Indeed, while 
metal-poor clusters in the outer halo and inner spheroid have comparable ages, 
there are apparent differences in photometric properties 
(Davidge, C\^{o}t\'{e}, \& Harris 1996, Davidge \& Courteau 1999).

	Although there is a strong scientific motivation to study globular 
clusters of all metallicities at low Galactic latitudes, efforts to use data 
obtained with CCD detectors to determine basic cluster 
properties such as distance, reddening, and metallicity 
have been frustrated by heavy, non-uniform line-of-sight extinction 
and contamination from disk and bulge stars. An added complication for 
metal-rich clusters is that line blanketing affects the photometric properties 
of the most evolved stars at visible wavelengths 
(e.g. Bica, Barbuy, \& Ortolani 1991).

	Observations at wavelengths longward of 1$\mu$m are less susceptible to 
these problems, as extinction decreases with increasing 
wavelength, while the effects of line blanketing are 
greatly reduced, making it easier to trace the CMDs of 
metal-rich clusters and identify features such as the tip of the first 
ascent giant branch (hereafter RGB-tip) (e.g. Davidge \& 
Simons 1994). This wavelength region also contains temperature and 
metallicity-sensitive molecular transitions, such as the 
$2.3\mu$m CO bands, the depths of which can be estimated from narrow-band 
images, thereby providing an efficient means of obtaining line 
strength information for large numbers of objects (e.g. Davidge 1998).

	In the current study, moderately deep near-infrared images are used to 
investigate four globular clusters at low Galactic latitudes: Liller 1, 
Djorgovski 1, HP 1, and NGC 6528. The first three were observed 
because they are located within a few degrees of the 
Galactic Center (GC) and, at least at the time of observation 
(1996), were thought to be metal-rich. As for NGC 6528, this cluster has been 
the target of many investigations and was observed to provide a comparison 
source for Liller 1. Various properties of these clusters, as listed in the May 
15 1997 version of the Harris (1996) database, are summarized in Table 1.

	Liller 1 is one of the most metal-rich globular clusters in the Galaxy, 
and so is an important target for studies of the chemical enrichment history of 
the inner spheroid. Armandroff \& Zinn (1988) concluded that [Fe/H] $\sim +0.2$
based on the strength of the near-infrared calcium triplet in the integrated 
cluster spectrum, but also pointed out that there are significant 
uncertainties in this measurement due to contamination from field stars and 
difficulties centering the cluster on the spectrograph 
slit. The depths of stellar absorption features near $1.6\mu$m 
in integrated cluster spectra suggest that Liller 1 has a metallicity 
similar to the clusters NGC 6553 and NGC 6528 (Origlia et al. 1997), which have 
roughly solar metallicities on the Zinn \& West (1984) scale.

	Frogel, Kuchinski, \& Tiede (1995) conclude that (1) the color and 
slope of the Liller 1 giant branch on the $(K, J-K)$ CMD are consistent with a 
metallicity that is solar or higher, and (2) the brightness and color of the 
HB imply a distance modulus $\mu_0 = 14.7 \pm 0.2$ 
with $E(J-K) = 1.7$, which corresponds to $E(B-V) = 3.3$ using the Rieke \& 
Lebofsky (1985) reddening law. For comparison, Armandroff \& Zinn (1988) 
concluded that $E(B-V) = 2.7$ based on the equivalent width of the $\lambda 
8621$ absorption feature. Ortolani, Bica, \& Barbuy (1997a) obtained deep 
$VRIz$ observations of Liller 1, and the CMDs constructed from these data 
reveal the curved upper giant branch characteristic of metal-rich systems. 
Ortolani et al. (1997a) suggest that Liller 1 is more metal-rich than NGC 6553 
and NGC 6528 based on the curvature of the upper giant branch; 
however, it is not clear what fraction of the stars in the Ortolani 
et al. (1997a) study are actual cluster members.

	Djorgovski 1, which was discovered by Djorgovski (1987) during a search 
for optical counterparts to IRAS sources, is the least studied cluster in the 
present sample. Djorgovski noted that this cluster is heavily reddened, and 
speculated that it may have a small Galactocentric distance. However, a 
subsequent investigation of the integrated properties of Djorgovski 1, 
combined with photometric measurements of the brightest stars, led 
Mallen-Ornelas \& Djorgovski (1993) to conclude that $E(B-V) = 1.7 \pm 0.3$ and 
$\mu_0 = 17.1 \pm 0.5$, placing this cluster on the far side of the 
Galaxy. Ortolani, Bica, \& Barbuy (1995) have since obtained deep 
$V$ and $I$ images of the cluster and surrounding field and, 
while agreeing with Mallen-Ornelas \& Djorgovski (1993) that $E(B-V) = 
1.7 \pm 0.1$, concluded that [Fe/H] $\sim -0.4$ and $\mu_0 = 14.72 \pm 0.15$ 
based on a small population of red stars presumably located on the 
upper giant branch.

	Early spectroscopic studies of HP 1 were suggestive of a moderately 
high metallicity. Armandroff \& Zinn (1988) concluded that [Fe/H] $= -0.56$ and 
$E(B-V) = 1.44$ based on the strengths of the near-infrared Ca triplet and the 
$\lambda 8621$ feature in the integrated cluster spectrum. Minitti (1995b) 
measured line strengths and near-infrared colors for six bright stars and 
concluded that [Fe/H] $\sim -0.3 \pm 0.2$ and $E(J-K) = 0.94 \pm 0.10$, which 
corresponds to $E(B-V) = 1.8 \pm 0.2$. However, photometric studies challenge 
these estimates. The reddening-corrected $(K, J-K)$ CMD of HP 1 
presented by Minniti, Olszewski, and Rieke (1995) shows a steep, relatively 
blue giant branch, indicative of a low metallicity. Ortolani, 
Bica, \& Barbuy (1997b) presented an $(I, V-I)$ CMD of HP 1 that shows a blue 
HB and a giant branch that is well matched by the metal-poor globular cluster 
NGC 6752, for which [Fe/H] = $-1.54$ (Zinn \& West 1984). Ortolani et al. 
(1997b) further conclude that $E(B-V) = 1.19$ and $\mu_0 = 14.2 \pm 
0.2$, and suggest that the earlier metallicity estimates for HP 1 were affected 
by contamination from bulge stars.

	In contrast to the other three clusters discussed in this paper, there 
is general agreement concerning the properties of NGC 6528. 
Optical CMDs of NGC 6528 published by van den Bergh \& Younger (1979), 
Ortolani, Bica, \& Barbuy (1992), and Richtler et al. (1998) reveal a giant 
branch that, although broadened by differential reddening and prone to 
significant contamination from field stars, shows the curvature that is 
one of the signature characteristics of metal-rich clusters. 
Zinn (1980) concludes that $E(B-V) = 0.56$, while 
Reed, Hesser, \& Shawl (1988) find that $E(B-V) = 0.73$ based on 
integrated cluster colors. Ortolani et al. (1992) conclude that $E(B-V) = 0.55$ 
based on the color of the HB.

	NGC 6528 has the largest near-infrared Ca triplet equivalent width in 
the sample compiled by Rutledge, Hesser, \& Stetson (1997), and 
Armandroff \& Zinn (1988) calculate that [Fe/H] $= -0.33$ using this 
measurement. Zinn \& West (1984) find that [Fe/H] $= +0.12$ from the Q$_{39}$ 
index, while Cohen \& Sleeper (1995) conclude that [Fe/H] $= +0.1$
based on the slope of the giant branch on the $(K, J-K)$ CMD. As noted 
previously, Origlia et al. (1997) find that NGC 6528 and Liller 1 have similar 
metallicities based on the strength of absorption features near $1.6\mu$m.

	The paper is organized as follows. The observations, which were 
obtained with the Cerro Tololo Interamerican Observatory (CTIO) 
CIRIM camera and the Canada-France-Hawaii Telescope (CFHT) 
adaptive optics (AO) system $+$ KIR camera, are discussed in \S 2, along with 
the procedures that were employed to reduce the data. 
The photometric measurements and the CMDs, luminosity functions (LFs), and 
two-color diagrams (TCDs) are presented in \S 3. The properties of the 
surrounding bulge fields and the clusters are investigated in 
\S 4 and \S 5. A summary and discussion of the results follows in \S 6.

\section{OBSERVATIONS \& REDUCTIONS}

\subsection{CTIO Data}

	Fields in and around Liller 1, Djorgovski 1, and HP 1 were observed 
during a five night observing run in July 1996 with the CTIO CIRIM camera, 
which was mounted at the Cassegrain focus of the 1.5 meter telescope. CIRIM 
contains a $256 \times 256$ Hg:Cd:Te array, with an image scale of 0.6 arcsec 
pix$^{-1}$, so that the detector samples a $154 \times 154$ arcsec field. A 
series of short exposures with each cluster centered on the detector (CTIO 
Field 1) were recorded through $J, H, Ks$, $2.2\mu$m continuum, and CO filters. 
Deeper $J, H,$ and $Ks$ images of areas 2 arcmin (CTIO Field 2) and 30 arcmin 
(CTIO Field 3) north of Liller 1 and HP 1 were also recorded to study the 
surrounding bulge population and monitor field star contamination. 

	Four exposures offset in a $5 \times 5$ arcsec dither pattern were 
recorded of each field. Additional details of the observations, including the 
co-ordinates of the field centers, exposure times, and image quality can be 
found in Table 2. Photometric standard stars from Casali \& Hawarden (1992) and 
Elias et al. (1982) were also observed throughout the observing run.

	The data were reduced using the procedures described by Davidge \& 
Courteau (1999), with corrections being applied for dark current, flat-field 
variations, time-dependent variations in the mean sky level, and 
thermal emission signatures. The reduced images for each field $+$ filter 
combination were spatially registered to correct for the offsets introduced 
during acquisition, and then median-combined to reject bad pixels and cosmic 
rays. The final $K$ images are shown in Figure 1.

\subsection{CFHT Data}

	Fields in Liller 1 and NGC 6528 were observed with the CFHT AO system 
(Rigaut et al. 1998) and KIR imager during the night of Sept 4 - 5 UT 1998. 
KIR contains a $1024 \times 1024$ Hg:Cd:Te array with an image scale of 0.034 
arcsec pix$^{-1}$, so that the imaged science field is $34.8 \times 34.8$ 
arcsec. The CFHT AO system uses stars as reference beacons for atmospheric 
compensation, and the large line-of-sight extinction towards Liller1 
renders the brightest members of this cluster too faint at 
visible wavelengths to serve as reference beacons. Therefore, the star 
GSC07380-00845, which is 1.5 arcmin from the cluster center, was used for this 
purpose. Deep $J, H,$ and $Ks$ images of two fields on the line connecting 
GSC07380-00845 with the center of Liller 1 were recorded; CFHT Field 1 is 
closest to the cluster, sampling distances 0.8 to 1.5 arcmin from the cluster 
center, while CFHT Field 2 samples an interval 1.5 to 2.3 arcmin from the 
cluster center.

	NGC 6528 is less heavily reddened than Liller 1, and the brightest 
cluster members can be used as reference sources, thus permitting the cluster 
center, where the density of cluster members is greatest, 
to be observed. $J, H,$ and $Ks$ images of the center of NGC 6528 were 
supplemented with images recorded through CO and $2.2\mu$m continuum filters. 

	Four exposures per filter, offset in a square dither pattern, were 
recorded of each field. A grid of photometric standard stars from Casali \& 
Hawarden (1992) and Elias et al. (1982) were also observed during the course of 
the three night observing run. Additional details of the observations can be 
found in Table 3.

	The data were reduced using the procedures described in \S 2.1, 
and the final $K$ images for each field are shown in Figure 2.

\section{RESULTS}

\subsection{Photometric measurements}

	A single point-spread function (PSF) was constructed for each 
field$+$filter combination using routines in 
the DAOPHOT (Stetson 1987) photometry package, and the brightnesses of 
individual stars were then measured with ALLSTAR 
(Stetson \& Harris 1988). Anisoplanaticism causes the PSF to vary with distance 
from the reference star in AO-compensated data, and this is a potential 
source of concern for the CFHT observations. However, the effects of 
adopting a single PSF on the photometric measurements are not substantial over 
the $35 \times 35$ arcsec KIR field (Davidge \& Courteau 1999), and this 
point is investigated further with the NGC 6528 data in the next section.

	Integrated near-infrared colors and CO indices were also measured for 
Djorgovski 1, HP 1, and Liller 1. The aperture sizes for these measurements 
were determined by finding where the surface brightness profiles for each 
cluster equaled that of the surrounding sky on the Field 1 $K$ images. 
The colors and CO indices measured in this manner are listed in Table 4, 
along with the aperture radii. In \S 5.5 it is shown that the 
integrated photometric properties of Djorgovski 1 are affected by bulge stars, 
which are the brightest objects in the cluster field, and the entries in 
brackets show the colors measured for this cluster with the brightest stars 
removed.

\subsection{NGC 6528}

	The $K$ LF of NGC 6528, shown in the top panel of Figure 3, indicates 
that stars as faint as $K \sim 17.5$ are detected with these data.
The HB, which occurs near $K \sim 13.2$, is the most conspicuous feature in the 
LF, while the RGB-bump, which forms when the hydrogen burning shell encounters 
the discontinuity in hydrogen content that defines the maximum extent of 
the convective envelope during main sequence evolution (e.g. Iben 1968), occurs 
at $K \sim 13.8$. The brightness and amplitude of the RGB-bump are 
metallicity-dependent, becoming stronger and fainter as metallicity increases. 
The brightness difference between the RGB-bump and the HB changes with 
metallicity among metal-poor clusters (e.g. Fusi Pecci et al. 1990), and the 
current data indicates that this trend continues to even higher 
metallicities, with the RGB-bump being easily detected at near-infrared 
wavelengths. In particular, Ferraro et al. (1994) found that the RGB-bump in 
the [Fe/H] $\sim -0.6$ (Zinn \& West 1984) globular cluster M69 is located 0.15 
mag in $K$ fainter than the HB, while the RGB-bump in NGC 6528, which has 
a roughly solar metallicity (\S 1), is $\sim 0.6$ mag fainter than the HB.
Consequently, the difference in brightness between the HB and RGB-bump provides 
an independent means of ranking the metallicities of heavily obscured 
metal-rich clusters. 

	The $K$ HB LF of NGC 6528 is compared in Figure 4 with the LFs of the 
moderately metal-rich ([Fe/H] $\sim -0.5$; Zinn \& West 1984) globular clusters 
NGC 6304 and NGC 6316, taken from Tables 10 and 11 of Davidge et al. 
(1992). The NGC 6304 and NGC 6316 LFs have been shifted along the 
brightness axis to match the peak in the NGC 6528 LF, and 
scaled to match the number of HB stars in NGC 6528. A Kolmogoroff-Smirnoff 
(K-S) test indicates that the LFs of all three clusters are not significantly 
different. 

	The $(H, J-H)$ and $(K, J-K)$ CMDs of NGC 6528 are shown in Figure 5. 
The scatter in the CMDs is $\pm 0.05$ mag in $J-H$ between $H = 14$ and 15, 
and $\pm 0.08$ mag in $J-K$ between $K = 13.5$ and 14.5. This relatively modest 
scatter, coupled with the excellent agreement between the HB LFs shown in 
Figure 4, strongly suggests that the use of a single PSF has not introduced a 
large amount of scatter into the photometric measurements.

\subsection{Liller 1}

	The CTIO Field 1 observations sample the central regions of the 
cluster, where the ratio of cluster to bulge stars changes rapidly with 
distance from the cluster center. The photometric properties 
of stars in two annuli in this field will be 
examined separately: annulus 1 extends from the cluster center 
to the radius defined for the aperture measurements (\S 3.1), while annulus 2 
extends from annulus 1 to the edge of the field. The central fields of 
Djorgovski 1 and HP 1 will be similarly divided.

	The $K$ LFs of the CFHT Liller 1 fields are shown in the lower panels 
of Figure 3, while the LFs for the CTIO Liller 1 fields are plotted 
in Figure 6. The number density of stars in annulus 1 
clearly exceeds that in annulus 2 at the bright end, as expected if 
the most luminous stars in Liller 1 are at least as bright 
as those in the surrounding bulge. Moreover, despite being recorded with 
significantly longer integration times, the data for CTIO Fields 2 and 3 do 
not go much fainter than the CTIO Field 1 observations, indicating that, 
with an image quality of 1.2 arcsec FWHM, photometric completeness
is controlled more by crowding than photon statistics. 

	The $(H, J-H)$ and $(K, J-K)$ CMDs of Liller 1 and the surrounding 
fields are shown in Figures 7 -- 10. The HB can be seen in the CFHT Field 1 CMD 
as a clump of stars with $H \sim 15.0$ and $K \sim 14.4$; the HB is also 
evident as a modest bump in the $K$ LFs of the CFHT data in Figure 3. 
Frogel et al. (1995) found that the brightest HB stars in 
Liller 1 have $K = 14.1$ and, given that the HBs of metal-rich 
clusters typically have widths of a few tenths of a magnitude, 
suggested that the HB in Liller 1 should occur near $K = 14.4$. 
The current data clearly confirm this prediction.

	Contamination from bulge and disk stars is significant in the CFHT 
data, which samples moderately large distances from the cluster 
center. The most conspicuous non-cluster component in the CMDs is the plume of 
blue disk stars, which merges with the giant branch near $H \sim 19$ and $K 
\sim 18.5$, making it difficult to determine the color and brightness of the 
Liller 1 main sequence turn-off from the CFHT data.

	The number density of disk stars is of interest 
for Galactic structure studies, and the LF of stars with $J-K \leq 
1.5$ in CFHT Fields 1 and 2 and CTIO Fields 2 and 3 is shown in 
Figure 11. The number densities of blue disk stars measured from the various 
datasets define a continuous smooth sequence, indicating that the spatial 
distribution of these objects is uniform in the vicinity of Liller 1. A 
linear least squares fit to the data in Figure 11 yields the relation:

\vspace{0.3cm}
\hspace*{2.0cm}log(n$^{blue}_{05}$) = ($0.354 \pm 0.019) \times K - (7.776 \pm 0.041$) 

\vspace{0.3cm}
\noindent{where} n$^{blue}_{05}$ is the number of blue stars per square arcsec 
per 0.5 mag interval. This relation is shown as a dashed line in Figure 11.

	The severity of disk and bulge star contamination on the giant branch 
of the CFHT CMDs can be estimated using star counts from CTIO Field 3, which 
contains no cluster stars and, based on the LF in Figure 6, is largely complete 
to $K \sim 14$. CFHT Fields 1 and 2 have 61 and 38 stars with $K$ between 11 
and 14 and $J-K \geq 1.5$, corresponding to densities of 0.0497 and 0.0310 
stars per square arcsec. For comparison, the density of stars in 
this brightness and color range in CTIO Field 3 is 0.0170 stars 
per square arcsec. The ratio of cluster to field stars in CFHT 
Field 1 is then (0.0497-0.0170)/0.0170 = 1.9:1, while in Field 2 it is only 
(0.0310 - 0.0170)/0.0170 = 0.8:1. Therefore, the giant branch of CFHT 
Field 1 is dominated by cluster stars, while CFHT Field 2 contains a 
roughly equal mix of cluster and bulge stars.

	There is a concentration of stars in CFHT Fields 1 and 2 near 
$K \sim 15.2$. While the relative brightness of these objects with respect to 
the Liller 1 HB indicates that they could belong to the RGB-bump, 
this is unlikely as they are uniformly distributed over both 
fields, suggesting membership in a background population. Indeed, 
there are $31 \pm 6$ objects with $J-K$ between 1.6 
and 2.2 and $K$ between 14.7 and 15.5 in CFHT Field 1, and $25 \pm 5$ objects 
in the corresponding area of the CFHT Field 2 CMD. These stars 
are too faint to belong to the bulge HB and, given that Liller 1 is less than a 
degree from the Galactic Plane, may belong to a background HB 
population, likely originating in the disk {\it behind} the bulge. 
If the distance modulus of Liller 1 is 14.7 (Frogel et al. 1995) 
and all obscuring material is located in front of Liller 1, then 
HB stars at $K \sim 15.2$ would have a distance modulus $\mu_0 = 
15.5$. These objects, if disk HB stars, are thus $\sim 4.5$ kpc behind the 
GC, placing them on the periphery of the bulge.

	CO indices provide another means of comparing the 
properties of stars in Liller 1 and the surrounding bulge. 
The $(K, CO)$ CMDs of CTIO Field 1 annuli 1 and 2 are compared in 
the top panels of Figure 12, while the histogram distributions 
of CO values for stars with $K \leq 12$, where photometric errors are modest, 
are compared in the lower panels. The CO distributions, which have been 
normalized to the total number of stars in each annulus with $K \leq 12$, 
are different in that the annulus 1 distribution is narrower and skewed 
towards lower CO values, and a K-S test indicates that the two 
distributions differ at the 97\% confidence level.
This result could indicate that (1) at a given 
brightness stars in Liller 1 have intrinsically weaker CO features than 
those in the bulge, (2) the reddening in annulus 1 is higher than that in 
annulus 2, and/or (3) the distance modulus of Liller 1 is smaller than that 
of the GC. The second possibility is ruled out because the annulus 
1 giant branch locus is not redder than that of annulus 2; indeed, the 
annulus 1 giant branch is actually slightly bluer than 
that of annulus 2 (\S 5). As for the third possibility, 
the distance modulus infered for Liller 1 by Frogel et al. (1995) suggests that 
the cluster is more distant than the GC, and accounting for this 
would only exacerbate the differences evident in the lower panel of Figure 12 
due to the luminosity dependence of the CO bands. Therefore, these data 
indicate that at a given brightness stars in Liller 1 and 
the surrounding bulge have intrinsically different CO strengths.

\subsection{Djorgovski 1}

	The $K$ LFs of annuli 1 and 2 in Djorgovski 1 Field 1 are shown in 
Figure 13. The number density of stars in annulus 1 exceeds that in annulus 2 
when $K \leq 12.5$, indicating that the most luminous stars in Djorgovski 1 are 
fainter than the brightest bulge giants. The $(H, J-H)$ and $(K, 
J-K)$ CMDs of annuli 1 and 2 are plotted in Figures 14 and 15. The 
concentrations of stars near $H \sim 14$ and $K \sim 13.5$ are due to the 
bulge HB, and there is a modest population of blue disk stars in both annuli.

	The CO index provides a way of distinguishing between stars in 
Djorgovski 1 and the surrounding bulge. The $(K, CO)$ 
CMDs of annuli 1 and 2 are plotted in the top 
panels of Figure 16, while the histogram distributions of CO 
indices for stars with $K \leq 13$, where photometric errors are relatively 
small, are compared in the lower panels. The CO distributions of annuli 1 and 
2, which have been normalized according to the number of stars with $K < 13$ in 
each annulus, have different appearances, in the sense that the annulus 1 
distribution is wider and peaks at lower values; a K-S test indicates that 
the two CO distributions differ in excess of the 99\% 
confidence level. The giant branch loci on the CMDs indicate that annuli 1 and 
2 are subjected to similar amounts of interstellar extinction, while there is 
no evidence to suggest that the distance modulus of Djorgovski 1 
is significantly smaller than that of the GC; in fact, Djorgovski 1 is 
likely located on the far side of the bulge (\S 5). Therefore, the comparison 
in the lower panel of Figure 16 indicates that stars in Djorgovski 1 have 
significantly smaller CO indices, and hence are more metal-poor, than stars in 
the surrounding bulge. 

\subsection{HP 1}

	The $K$ LFs of the HP 1 fields are shown in Figure 17. There is an 
excess of stars in annulus 1 with respect to annulus 2 when $K \geq 10$, 
indicating that HP 1 contains stars spanning almost the same range of 
brightnesses as those in the bulge. The LFs of annulus 2, Field 2, and Field 3 
are similar at the bright end, and the bump centered at $K = 13.25$ in the 
Field 2 and 3 LFs is due to the bulge HB.

	The $(H, J-H)$ and $(K, J-K)$ CMDs of the HP 1 fields are compared in 
Figures 18 and 19. The tight CMD sequences indicate that the effects of 
differential reddening are modest along this line of sight. The annulus 
1 CMDs show a blue population that is not prevalent in annulus 2, 
and this is most evident at the bright end, near $K \sim 10$ and $H \sim 11$.

	The annuli 1 and 2 $(K, CO)$ CMDs are compared in the upper panels of 
Figure 20, while the histogram distributions of CO indices for stars with $K 
\leq 12$ are shown in the lower panels. A K-S test indicates that the CO 
distributions differ at only the 85\% confidence level, even though the 
LFs indicate that there is a significant number of cluster stars in annulus 1 
when $K \geq 10$. While this comparison provides only weak evidence that the CO 
distributions of stars in HP 1 and the bulge are different, the mean CO values 
in the two annuli differ by $0.039 \pm 0.014$ mag, in the sense that the CO 
index in annulus 2 is higher. Moreover, when the annulus 2 distribution is 
subtracted from the annulus 1 distribution a significant excess of stars 
remains in annulus 1 at low CO values. Hence, it is concluded that 
stars in HP 1 and the surrounding bulge have different CO indices. 

\section{REDDENING ESTIMATES AND FIELD STAR PROPERTIES}

\subsection{Reddening Estimates}

	The photometric properties of the brightest, most metal-rich, stars in 
the inner bulge do not change with radius (Davidge 1998), and so reddening 
estimates for Liller 1, Djorgovski 1, and HP 1 can be obtained by comparing the 
colors of stars in the fields surrounding each cluster with those of giants in 
BW. To facilitate these comparisons, normal points were generated by computing 
the mean $J-H$ and $J-K$ colors in $\pm 0.25$ mag intervals along the $H$ and 
$K$ axes of the annulus 2 $(H, J-H)$ and $(K, J-K)$ CMDs, and the results are 
shown in Figures 21 and 22; the Field 2 and 3 data for Liller 1 and HP 1 were 
not used to estimate reddenings because the brightest stars are saturated. The 
BW M giant sequence listed in Table 3B of Frogel \& Whitford (1987), plotted as 
open squares in these figures, was then shifted along the Rieke \& Lebofsky 
(1985) reddening vector to match the normal points for each cluster field. 

	The color excesses that give the best agreement between 
the BW and cluster field sequences are shown at the top of each panel in 
Figures 21 and 22, while the corresponding $E(B-V)$ values 
are listed in the second and third columns of Table 5. The 
BW and field star sequences can be matched to within 
$\pm 0.02$ mag. Given that there is a 
$\pm 0.03$ mag uncertainty in the color calibration, 
then the uncertainty in the near-infrared color excesses is $\pm 0.04$ 
mag. Hence, the individual $E(B-V)$ entries in Table 5 have an 
uncertainty of $\pm 0.1$ mag, not including possible systematic errors in 
the reddening curve. The two $E(B-V)$ values derived for 
each cluster agree to within this uncertainty, and so each pair was averaged, 
with the results listed in the last column of Table 5. 

	If the brightest stars in BW and the fields surrounding 
the clusters have similar photometric properties then they should follow 
similar $K$ versus CO relations. To see if this is the case, normal points were 
computed from the annulus 2 $(K, CO)$ CMDs by calculating the mean CO index 
in $\pm 0.25$ mag bins along the $K$ axis. The $K$ versus CO relation 
for bright BW giants, projected along a reddening vector 
with $E(CO)/E(B-V) = -0.04$ (Elias, Frogel, \& Humphreys 1985) using the 
$E(B-V)$ values given in the last column of Table 5, 
is compared with the annulus 2 data in Figure 23.

	The uncertainty in the photometric zeropoint of the CO indices is $\pm 
0.03$ mag, and the Liller 1 field star and BW sequences on the ($K, CO)$ 
CMD agree to within this error at all brightnesses. While 
there is good agreement between the BW and field star sequences 
for Djorgovski 1 and HP 1 when $K \leq 10$, at 
fainter values the difference between the BW and bulge sequences becomes 
significant. However, it should be recalled that although annulus 2 is 
dominated by bulge stars, this area also contains cluster stars which, in 
the case of Djorgovski 1 and HP 1, have 
weaker CO indices than most bulge stars (\S 3). 
The departure of the Djorgovski 1 and HP 1 field star sequences from the 
locus of BW giants when $K \geq 10$ on the $(K, CO)$ CMD is likely due to 
metal-poor cluster stars skewing the mean CO indices to lower values than 
would occur if a pure bulge component were present. This bias is not evident 
when $K \leq 10$ as the brightest stars in HP 1 and Djorgovski 1 are 
fainter than the brightest bulge stars (\S 5). 

\subsection{The Bulge LFs}

	The M$_K$ LFs for Djorgovski 1 annulus 2, HP 1 Field 3, and Liller 1 
Field 3 are compared in Figure 24. Absolute brightnesses 
were computed using the $E(B-V)$ entries in the last column 
of Table 5, with the distance modulus of the GC set at 
$\mu_0 = 14.5$ (Reid 1993). The discontinuity in HP 1 Field 3 
when M$_K \leq -4$ is due to saturation of the brightest stars in this field.

	The field star LFs are parallel when $-4 \leq$ M$_K \leq -2$, and the 
HB is clearly evident in the HP 1 Field 3 data when M$_K \sim -1.5$. The 
HB is not conspicuous in the Djorgovski 1 LF due to contamination from cluster 
stars (recall that a separate background field was not observed for this 
cluster), while incompleteness becomes significant at the brightness of the 
HB in the Liller 1 data.

	The ratio of HB stars to bright RGB and AGB stars, R', is sensitive to 
helium abundance (e.g. Buzzoni et al. 1983). Minniti (1995c) 
measured R' in five bulge fields, and the mean of these values is 
$\overline{R'} = 1.48 \pm 0.09$, where the uncertainty is the error in the 
mean. Minniti (1995c) did not find evidence for a gradient in R', although the 
number of fields studied was modest, and the uncertainties in individual R' 
determinations is significant. HP 1 is of potential interest as it has a 
smaller projected distance from the GC than any of the fields studied by 
Minniti (1995c).

	Following the procedure described by 
Davidge (1991) and Minniti (1995c), a linear least squares 
fit was applied to the HP 1 Field3 $K$ LF above and below the HB 
to determine the relation between log(n$_{05}$) for first ascent giants and $K$ 
so that the number of giant branch stars in the luminosity bin containing the 
HB could be estimated. The upper giant branch contains a 
mixture of AGB and first ascent giants, while the lower giant branch contains 
only first ascent giants, so the number counts of the former must be 
corrected for AGB contamination prior to performing the least squares 
fit. This was done statistically, as bright AGB and RGB stars have 
similar near-infrared colors, making it difficult 
to distinguish between these objects on the CMDs. The relative 
time scales of RGB and AGB evolution near the RGB-tip are such that 20\% of 
the bright giants are likely on the AGB (Caputo et al. 1989), and 
so the LF bins brighter than the HB were multiplied by 0.80 prior to performing 
the least squares fit to account for the presence of AGB stars.

	The current data indicate that R' $\sim 1.14 \pm 0.11$ in 
HP 1 Field 3. While this is lower than the mean R' for the bulge, it is in 
excellent agreement with R' measured for field F589 by Minniti (1995c). This 
measurement is thus consistent with the conclusion reached by Minniti (1995c) 
that R' does not vary with radius in the Galactic bulge. 

\section{CLUSTER PROPERTIES}

\subsection{R' in NGC 6528 and other metal-rich clusters}

	To date, R' has not been measured in clusters as metal-rich as NGC 
6528. Applying the procedure described by Davidge (1991) and Minniti (1995c) to 
the NGC 6528 $K$ data indicates that R' $\sim 1.25 \pm 0.25$. 
This is consistent with R' measurements in moderately metal-rich globular 
clusters such as 47 Tuc (Buzzoni \& Fusi Pecci 1983) and M69 (Ferraro et al. 
1994), although the uncertainties in individual R' values are large. 
The mean R' for 47 Tuc, M69, and NGC 6528 is $\overline{R'} = 
1.27 \pm 0.09$, where the error is the uncertainty in the mean. This 
differs from the mean derived from the Minniti (1995c) bulge measurements at 
only the 2.3-$\sigma$ significance level, so the existing data do not provide 
convincing evidence for a difference in helium contents between stars in 
metal-rich (ie. [M/H] $\geq -1$) clusters and the bulge.

\subsection{The relative metallicities of Liller 1 and NGC 6528}

	In \S 3.3 it was shown that the histogram distribution of CO indices 
for bright stars in Liller 1 annuli 1 and 2 were significantly different, in 
the sense that stars in annulus 2 have larger CO indices on average. 
Nevertheless, the difference between the peak CO values is modest, 
and there is significant overlap between the two CO distributions. 
Therefore, rather than use CO indices to identify stars in Liller 1, 
it is assumed that annulus 1 is dominated by stars in Liller 1, while 
annulus 2 is dominated by bulge stars. 

	Normal points were created from the 
Liller 1 annuli 1 and 2 $(K, J-K)$, $(K, H-K)$, and $(K, CO)$ CMDs 
by computing the mean $J-K$, $H-K$, and CO indices in $\pm 0.25$ mag bins in 
$K$. The annuli 1 and 2 normal point sequences on the $(K, J-K)$ and $(K, CO)$ 
CMDs, de-reddened assuming that $E(B-V) = 3.13$, are plotted in the top 
panels of Figure 25. The bulge sequence has been shifted along the $K$ axis 
assuming that Liller 1 is more distant than the GC, with $\Delta 
\mu = 14.7$ -- 14.5 = 0.2 mag, based on the distance estimates of Frogel et al. 
(1995) for Liller 1 and Reid (1993) for the GC. 

	It is evident from Figure 25 that at a given $K_0$ stars in 
Liller 1 tend to have smaller CO indices than stars in the surrounding 
bulge, with $\Delta CO_0 = 0.033 \pm 0.015$ when $K_0 \leq 10.5$. Moreover, 
the $J-K$ colors of the Liller 1 and bulge data generally differ by $0.1 - 0.2$ 
mag, in the sense that Liller 1 has smaller $J-K$ values, although there is 
significant scatter in the bulge normal point sequence. These comparisons thus 
indicate that stars in Liller 1 are not as metal-rich as those in the 
surrounding bulge. 

	Normal points were also computed from the NGC 6528 observations 
and the results, de-reddened assuming that $E(B-V) = 0.56$ and shifted 
along the $K$ axis to register the HB brightness with that of Liller 1, are 
also shown in Figure 25. The Liller 1 and NGC 6528 sequences are in good 
agreement on the upper part of both CMDs when $K_0 \leq 9.5$, but when $K_0 
\geq 9.5$ the NGC 6528 and Liller 1 sequences diverge, in the sense that the 
bulge and NGC 6528 sequences overlap. Therefore, the metallicity of Liller 1 is 
comparable to, or perhaps even slightly lower than, that of NGC 6528.

	The $(CO_0, (J-K)_0)$ TCD, shown in the lower right panel of Figure 25, 
indicates that stars in Liller 1, NGC 6528, and the bulge follow the relation 
between CO and $J-K$ for solar neighborhood giants plotted in Figure 5 of 
Frogel \& Whitford (1987). For comparison, the sequence for the moderately 
metal-rich cluster 47 Tuc, derived from the data tabulated by Frogel, Persson, 
\& Cohen (1981) and also shown in Figure 25, falls well 
below the solar neighborhood data. While the CO index depends 
on parameters such as [O/Fe] and [C/Fe], the most likely interpretation 
of these results is that Liller 1 and NGC 6528 have near-solar metallicities. 

	The comparisons in the top panels of Figure 25 
indicate that stars in Liller 1 and NGC 6528 may be slightly more metal-poor 
than the brightest stars in the bulge; however, stars in these clusters and 
the bulge have similar near-infrared spectral energy distributions, 
possibly suggesting a shared chemical enrichment heritage that is different 
from other parts of the Galaxy. The near-infrared spectral-energy 
distributions of these clusters and the bulge are compared in Figure 26, 
which shows the $(J-H, H-K)$ TCD. It is evident that the normal points for NGC 
6528, Liller 1 and the bulge fall below the sequences for solar neighborhood 
giants defined by Bessell \& Brett (1988) and giants in 
47 Tuc, as derived from the observations of Frogel, Persson, \& Cohen (1981). 

\subsection{The metallicity and distance of Djorgovski 1}

	In \S 3 it was shown that at a given brightness stars in Djorgovski 1 
have weaker CO indices than bulge stars. The difference between the peaks of 
the annuli 1 and 2 CO distributions is much larger than what was seen for 
Liller 1, and so CO indices are used to identify stars in Djorgovski 1. To 
establish a simple criterion for identifying cluster members, the CO histogram 
distributions for annuli 1 and 2 were scaled to account for differences in 
areal coverage and then subtracted. A statistically 
significant excess in annulus 1 with respect to annulus 2 occured 
when CO $\leq 0.025$; therefore, stars with CO $\leq 0.025$ are assumed 
to belong to Djorgovski 1. This simple approach does not take into account 
photometric errors and the luminosity-dependence of CO indices, so the CMDs 
constructed for Djorgovski 1 using this procedure will be most reliable 
for the brightest stars, which were used to construct the CO distribution 
and where the photometric errors are smallest. The bulge 
also contains a metal-poor component (McWilliam \& Rich 1994) which will not 
be rejected with this criterion; however, contamination from metal-poor 
bulge stars should be modest given the localised high density of cluster stars.

	The $(K, J-K)$ CMD of stars with $CO \leq 0.025$, de-reddened assuming 
that $E(B-V) = 1.44$ (\S 4), is shown in the left hand panel of Figure 27. 
Also shown is the NGC 6528 normal point sequence and aperture measurements 
from Cohen, Frogel, \& Persson (1978) for the metal-poor halo 
clusters M13 ([Fe/H] $= -1.6$) and M92 ([Fe/H] $= -2.2$). 
The data for these clusters were registered along the vertical axis such that 
the brightest stars matched those in Djorgovski 1.

	The Djorgovski 1 and M92 sequences are very similar, while the M13 and 
NGC 6528 giant branches fall to the right of the Djorgovski 1 data, 
suggesting that Djorgovski 1 is extremely metal-poor. The difference 
in distance moduli between Djorgovski 1 and M92 infered from the 
brightest stars is $\Delta ^{M92} _{Djorg 1} \sim 0.8$. If $\mu_0 = 14.6$ 
for M92 (Harris 1996) then $\mu_0 = 15.4$ (ie. r $\sim 12.5$ kpc), 
indicating that while Djorgovski 1 is located on the far side of the bulge, it 
nevertheless falls within the solar circle. 

\subsection{The metallicity and distance of HP 1}

	Stars in HP 1 were identified using the procedure 
described in \S 5.3. When the CO distributions shown in 
the lower panel of Figure 20 are subtracted after adjusting for differences 
in field coverage, a statistically significant excess 
occurs in the annulus 1 residuals when CO $\leq 0.055$; hence, 
stars with CO $\leq 0.055$ are assumed to belong to HP 1.

	The $(K, J-K)$ CMD for HP 1, de-reddened assuming that 
$E(B-V) = 0.74$ (\S 4), is shown in the right hand panel of Figure 27. 
The giant branch loci of the clusters NGC 6528, M13, and M92, 
registered along the vertical axis using the brightest stars, are 
also shown. The HP 1 and M13 giant branches are in good agreement, while the 
NGC 6528 and M92 giant branches are offset to the right and left of the HP 1 
data. This comparison thus suggests that [Fe/H] $= -1.6$ for HP 1 on 
the Zinn \& West scale, which is consistent with the good agreement found by 
Ortolani et al. (1997) between the $(I, V-I)$ CMDs of HP 1 and the [Fe/H] 
$= -1.54$ (Zinn \& West 1984) cluster NGC 6752. The difference in distance 
moduli between HP 1 and M13 infered from the RGB-tip is $\Delta \mu^{M13} _{HP 
1} \sim 0.4$, so if $\mu_0 = 14.3$ for M13 (Harris 1996) then $\mu_0 = 14.7$. 

\subsection{De-reddened Cluster Colors and CO Indices}

	Aaronson et al. (1978) investigated the 
integrated near-infrared photometric properties of globular clusters, 
and established empirical relations involving 
metallicity. Relationships of this nature, plotted in Figure 1 of 
Aaronson et al. (1978), can be used to check the 
metallicities derived in the preceding sub-sections. 

	The de-reddened integrated colors of Liller 1, Djorgovski 1, and HP 1 
are listed in Table 6. The positions of these clusters on the $(J-K$, [M/H]) 
and (CO, [M/H]) diagrams are shown in Figure 28, where the solid line is 
the least squares fit from Figure 1 of Aaronson et al. (1978), shifted to 
lower metallicities by 0.1 dex based on the mean difference between the 
metallicity values for the calibrating clusters listed in Table 1 of Aaronson 
et al. (1978) and Table 6 of Zinn \& West (1984), 
while the dashed lines show the scatter envelope in the 
data considered by Aaronson et al. (1978).

	The $J-K$ color and CO index of Liller 1 are consistent 
with this cluster being metal-rich, while the colors and CO index of 
HP 1 are consistent with a low metallicity. However, the 
$J-K$ color and CO index of Djorgovski 1 are ostensibly inconsistent 
with a low metallicity. When interpreting this result it should be 
recalled that Djorgovski 1 is the most distant 
cluster in this sample, and so the integrated photometric properties 
are most susceptible to contamination from bright bulge stars. 
In an effort to determine if the presence of a small number of 
bright bulge stars close to the cluster center have affected the 
aperture measurements, all stars brighter than $K \sim 10.5$, which is the 
point at which stars in Djorgovski 1 appear in significant numbers, were 
subtracted from the images. The near-infrared colors and CO index 
measured after removing bright stars are shown in brackets in Tables 4 and 6, 
and are plotted as filled triangles in Figure 28. 
It is evident that the removal of bright field stars has a significant impact 
on the $J-K$ and CO values, and the revised measurements are 
consistent with Djorgovski 1 being metal-poor.

\section{DISCUSSION \& SUMMARY}

	Infrared images have been used to determine the reddenings, 
metallicities, and distances of the heavily obscured low-latitude globular 
clusters Liller 1, Djorgovski 1, and HP 1. The projected surface 
density of bulge stars is significant near all three clusters, and different 
techniques have been used to extract samples of likely cluster members.

	In the case of Liller 1, which is metal-rich with a high central 
density, stars within 45 arcsec of the cluster center are assumed to be 
cluster members. The colors and CO indices of stars selected in this manner 
indicate that Liller 1 is not as metal-rich as the brightest stars 
in the bulge, and this is consistent with the tendency for
globular clusters in external galaxies to be more metal-poor than 
the underlying star light (e.g. Forbes et al. 1996, Da Costa \& Mould 1988). 
The integrated visible spectrum of BW is suggestive of an approximately solar 
mean metallicity (Idiart, de Freitas Pacheco, \& Costa 1996), which agrees 
with the mean [Fe/H] for bulge K giants measured by McWilliam \& Rich (1994). 
Thus, Liller 1 might be expected to have a solar or lower metallicity. 

	The current data indicate that Liller 1 
has a metallicity similar to, and perhaps even slightly lower than, that of 
NGC 6528, in agreement with the strengths of near-infrared absorption features 
measured in the integrated spectra of both clusters (Origlia et al. 1997).
While the metallicity of NGC 6528 is not well 
established, it likely falls in the range -0.3 (Armandroff \& Zinn 1988) to 
$+0.1$ (Zinn \& West 1984). Hence, the metallicity of Liller 1 is likely 
solar or lower, as expected if it has a metallicity 
lower than that of stars in the surrounding field. If Liller 1 has a 
metallicity comparable to NGC 6528 then the RGB-bump should occur roughly 0.6 
mag in $K$ fainter than the HB, at $K \sim 15$. High angular-resolution 
near-infrared observations of stars in the central regions of Liller 1 will 
be able to verify this prediction.

	The $(J-H, H-K)$ TCD indicates that stars in NGC 6528 and Liller 1 have 
photometric characteristics that are indicative of the bulge, rather than the 
disk or outer halo, lending support to the suggestion made by Minniti (1995a) 
that metal-rich clusters formed as part of the bulge. McWilliam \& Rich (1994) 
conclude that `the enrichment pattern of bulge giants sets them clearly apart 
from disk giants of the same [Fe/H]'. Detailed abundance analyses 
of giants in metal-rich clusters like Liller 1 and NGC 6528 should 
provide an independent means of determining if low latitude metal-rich 
clusters formed from material that experienced a chemical enrichment 
history related to bulge stars.

	CO indices, derived from narrow-band images, have been 
used to identify stars in Djorgovski 1 and HP 1. 
The CO index does not provide a foolproof means of identifying 
cluster members since photometric errors, the surface gravity 
and temperature sensitivity of the CO bands, and the presence of metal-poor 
stars in the bulge will blur the boundary between bulge and cluster 
stars. Nevertheless, the integrated near-infrared 
properties of Djorgovski 1 and HP 1 are consistent with the 
metallicities deduced from the CMDs, suggesting that the samples 
selected using the CO index are dominated by cluster members.

	Djorgovski 1 and HP 1 are much more metal-poor than Liller 1. The 
color and slope of the Djorgovski 1 giant branch on near-infrared CMDs
suggests that [Fe/H] $\leq -2$. The distance modulus derived for Djorgovski 1 
from the brightest stars, which places this cluster on the far side of the 
bulge, but still within the inner Galaxy, is intermediate between the values 
advocated by Mallen-Ornelas \& Djorgovski (1993) and Ortolani et al. (1995). 
The Djorgovski 1 CMDs constructed by Ortolani et al. (1995) show considerable 
scatter, with no obvious cluster sequence. Given the high density 
of field stars in the vicinity of Djorgovski 1, and the relative faintness 
of cluster members, it seems likely that the 
Ortolani et al. (1995) CMD is dominated by bulge stars.

	The data presented in this paper suggest that 
[Fe/H] $\sim -1.6$ for HP 1, in agreement with the study of 
Ortolani et al. (1997). Previous spectroscopic studies have suggested that HP 1 
is metal-rich, and it seems likely that these conclusions were based on 
observations of field stars, which are the brightest objects in the vicinity of 
HP 1. The current work suggests that HP 1 may be located slightly beyond 
the GC, whereas Ortolani et al. (1997) place HP 1 
in front of the bulge. This difference in distance estimates is 
due to the adopted reddenings: Ortolani et al. (1997) assumed that $E(B-V) = 
1.2$, which is significantly higher than the $E(B-V) = 0.7$ infered here from 
bright field giants.

	Djorgovski 1 and HP 1 are important targets for future study 
because they are part of the metal-poor inner spheroid. 
Djorgovski 1 is of particular interest because, with [Fe/H] $\leq -2$, it is 
one of the most metal-poor clusters in the inner Galaxy. Field 
star contamination could be reduced in future studies by 
observing the centers of each cluster, where the density of 
cluster members is highest. Images with FWHM $\leq 0.2$ 
arcsec should be capable of identifying main sequence turn-off 
stars in these fields (e.g. Davidge \& Courteau 1999).

\pagebreak[4]
\begin{center}
FIGURE CAPTIONS
\end{center}

\figcaption
[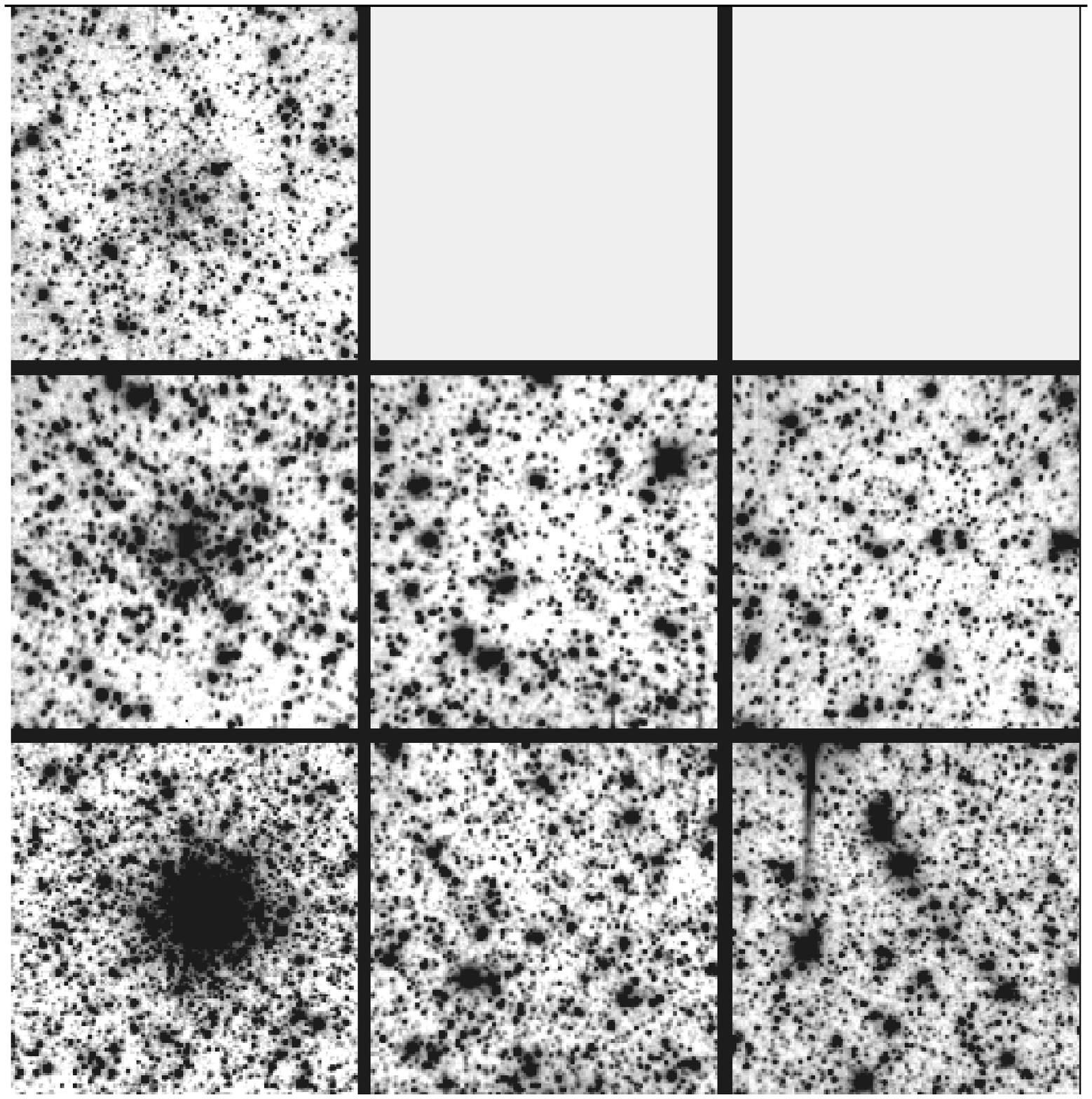]
{Final $K$ images for Djorgovski 1 (top row), HP 1 (middle row), and Liller 1 
(bottom row) recorded with the CTIO CIRIM camera. Images of, from left to 
right, CTIO Fields 1, 2, and 3 are shown for each cluster. Note that only 
one field, centered on the cluster, was observed for Djorgovski 1. Each image 
covers $2.5 \times 2.5$ arcmin. North is at the top, and East is to the left.}

\figcaption
[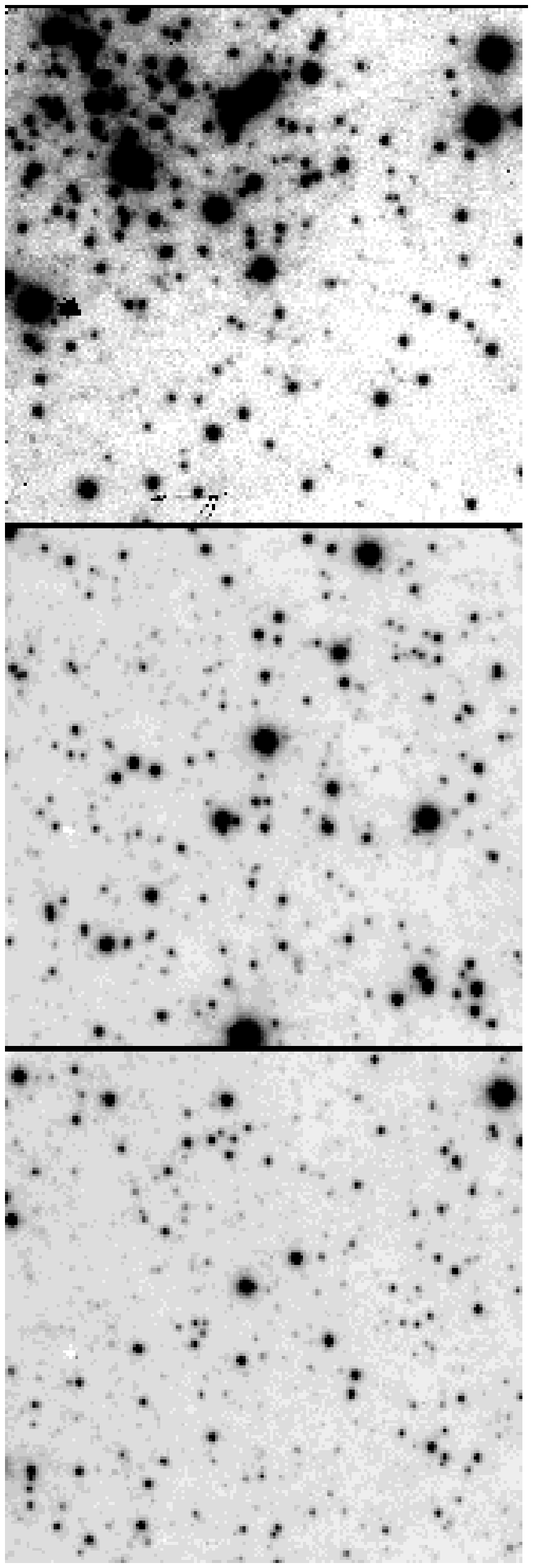]
{Final $K$ images for NGC 6528 (top panel), Liller 1 CFHT Field 1 
(middle panel), and Liller 1 CFHT Field 2 (bottom panel) recorded with the CFHT 
AOB+KIR. Each image covers $34 \times 34$ arcsec. North is at the top, 
and East is to the left.}

\figcaption
[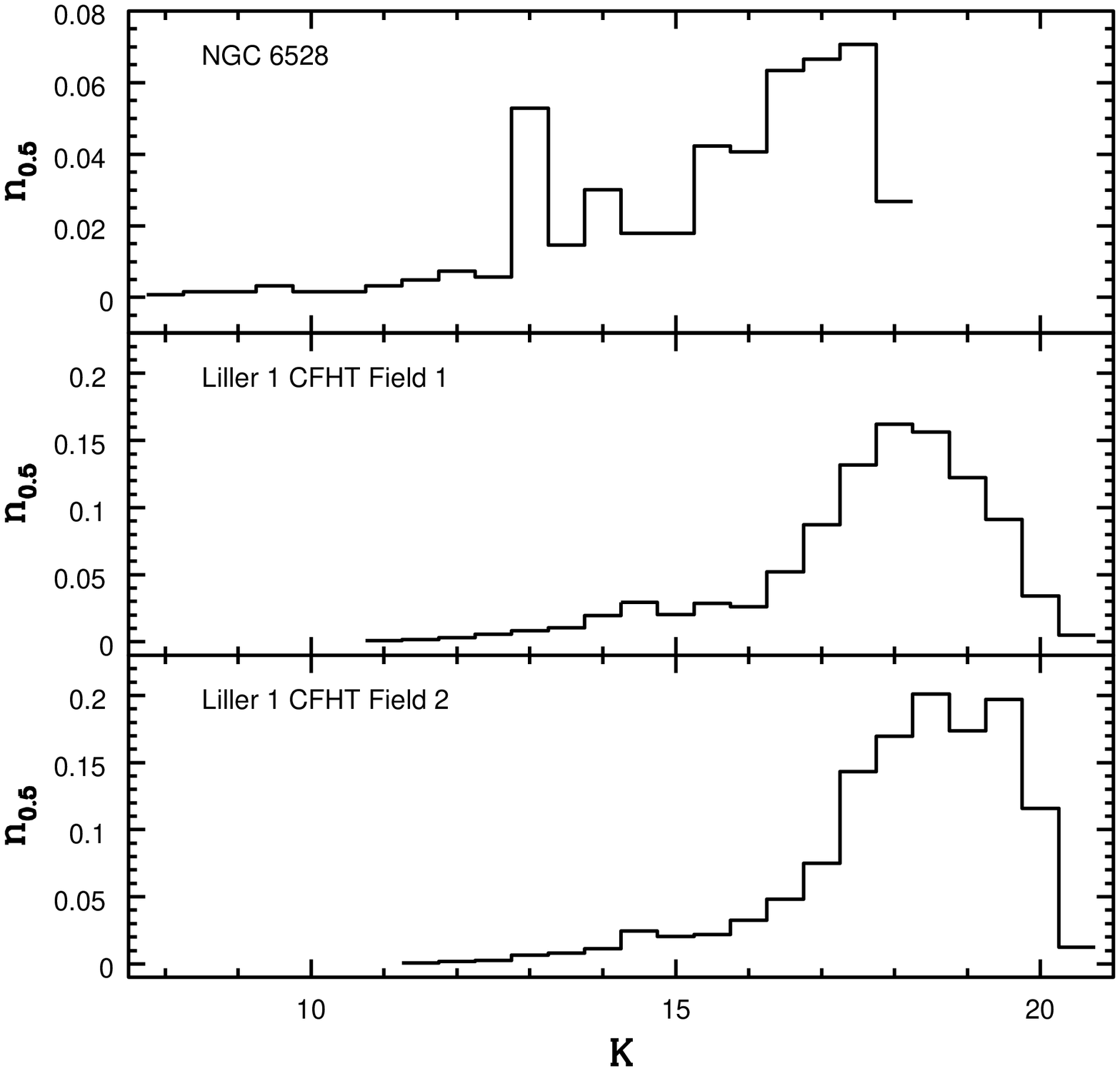]
{The $K$ LFs of NGC 6528 and Liller 1 obtained with the 
CFHT AOB+KIR. n$_{0.5}$ is the number of stars per 0.5 mag interval per 
square arcsec. The two peaks in the NGC 6528 LF are due to the HB ($K = 13.2$) 
and the RGB-bump ($K = 13.8$). The HB of Liller 1 is seen at $K = 14.5$.}

\figcaption
[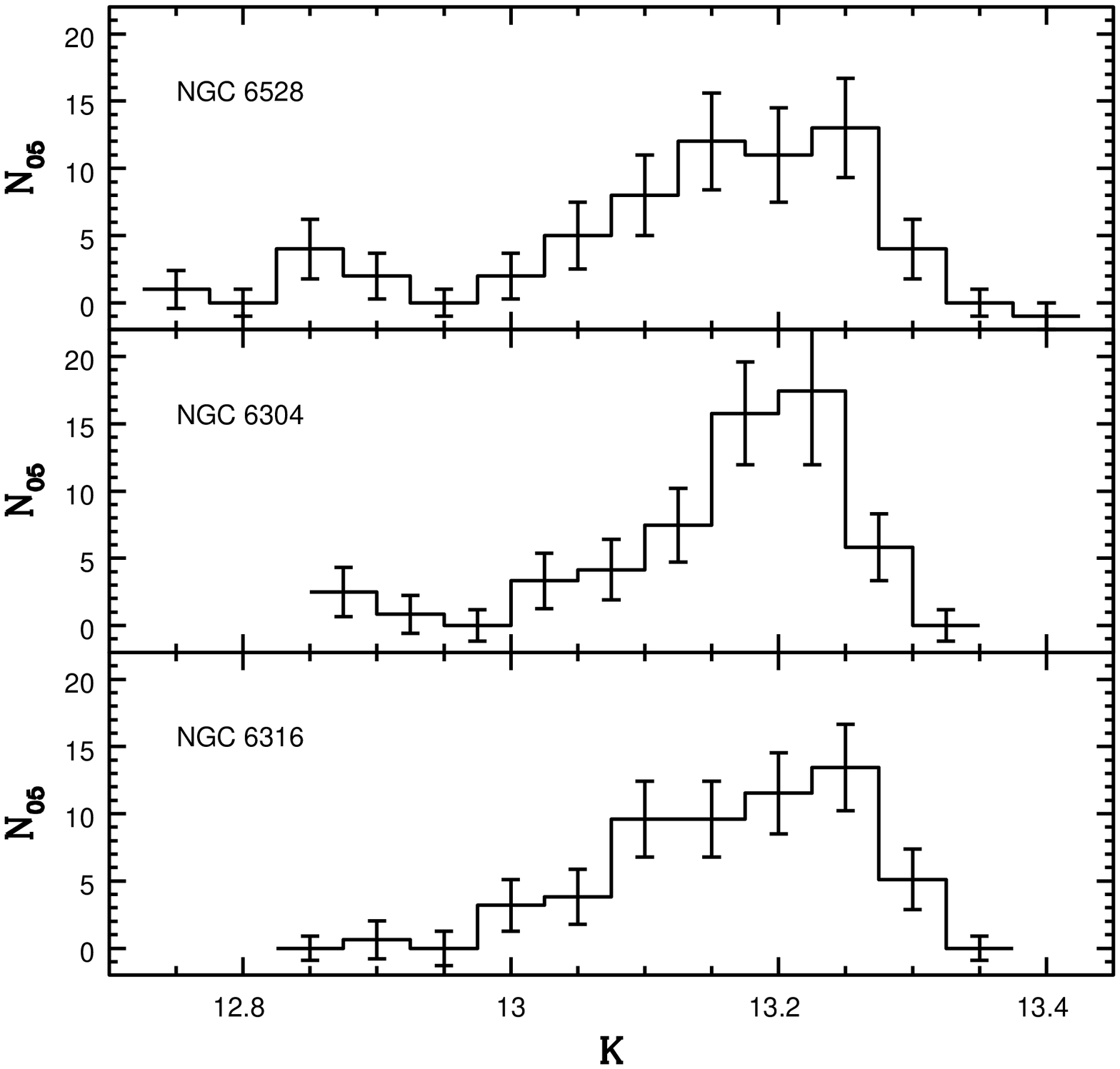]
{The $K$ HB LFs of NGC 6528, NGC 6304, and NGC 6316. The NGC 6304 and NGC 6316 
LFs are from Tables 10 and 11 of Davidge et al. (1992), and have been shifted 
along the $K$ axis to match the peak of the NGC 6528 LF after being 
scaled to match the number of stars in the NGC 6528 LF. N$_{0.5}$ is the 
number of stars per 0.5 mag interval. The errorbars show the uncertainties 
introduced by counting statistics.}

\figcaption
[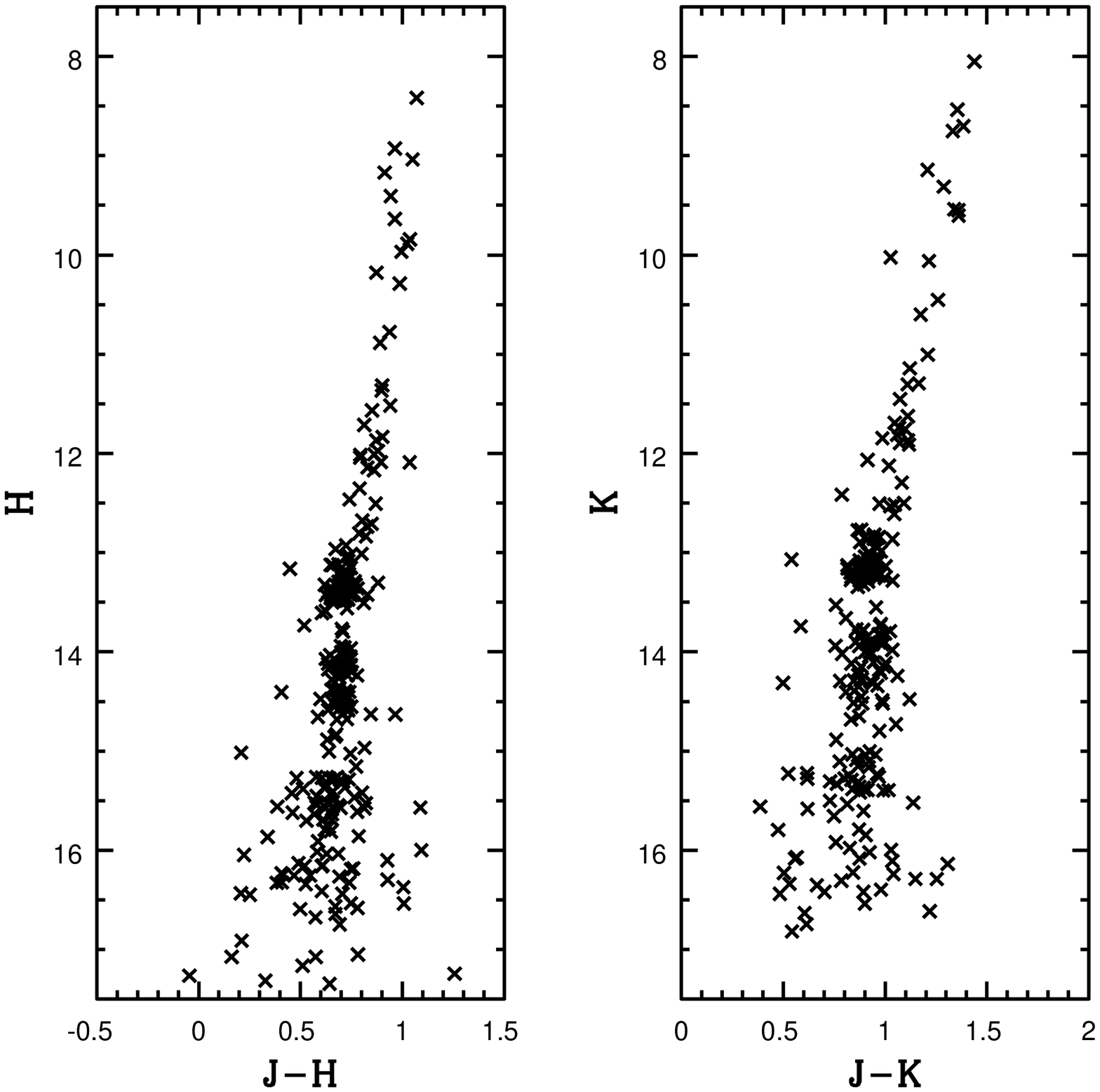]
{The $(H, J-H)$ and $(K, J-K)$ CMDs of NGC 6528. The HB is the clump of stars 
near $H \sim 13.4$ and $K \sim 13.2$}

\figcaption
[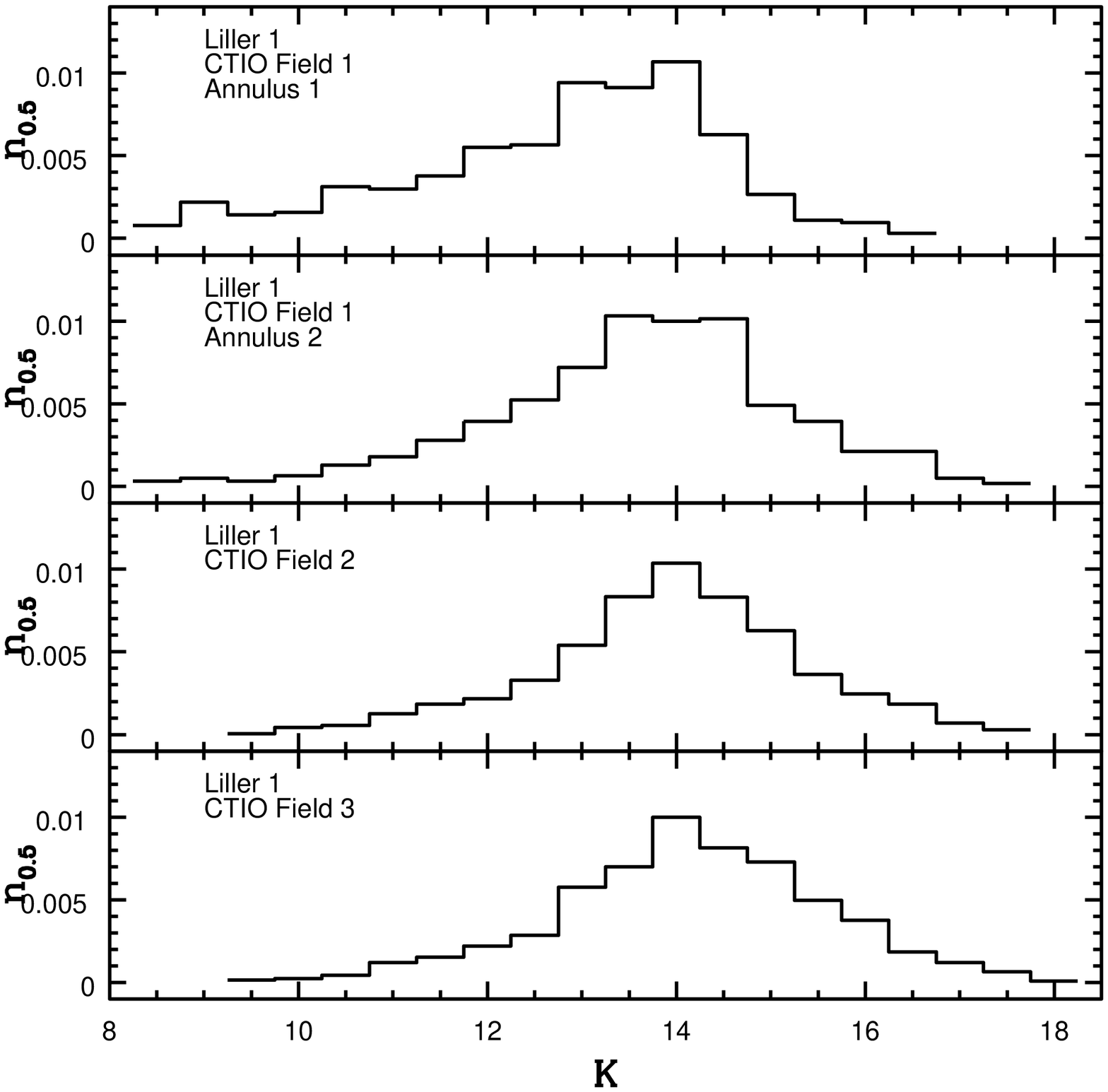]
{The $K$ LFs of the various Liller 1 fields observed with 
CIRIM at CTIO. n$_{0.5}$ is the number of stars per 0.5 mag interval per 
square arcsec. Field 1 has been divided into two annuli centered on the 
cluster. Note that incompleteness becomes significant 
at comparable brightnesses in the various fields, despite differences 
in integration time, indicating that the data are crowding-limited.}

\figcaption
[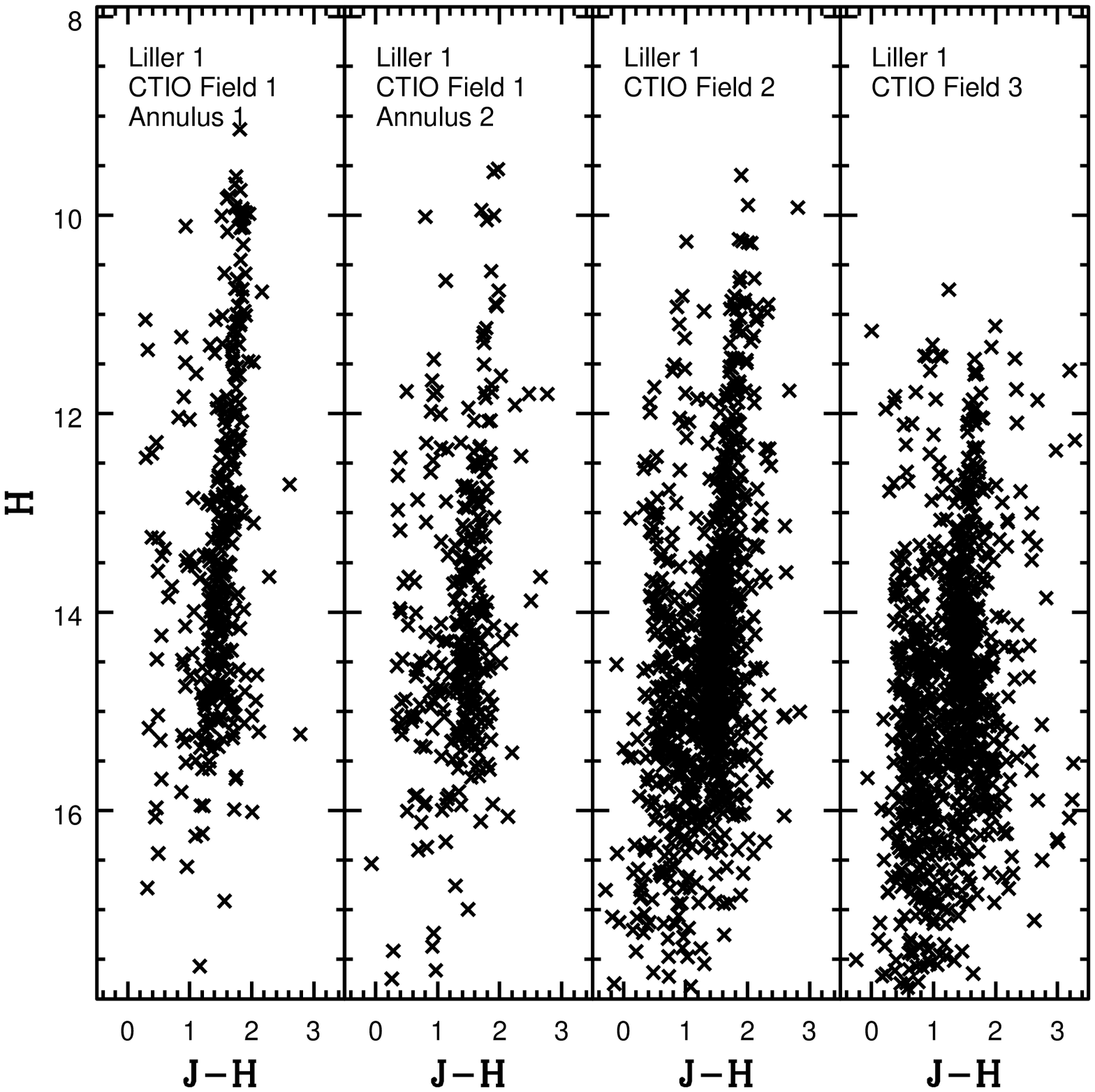]
{The $(H, J-H)$ CMDs of the Liller 1 fields observed with CIRIM at CTIO.}

\figcaption
[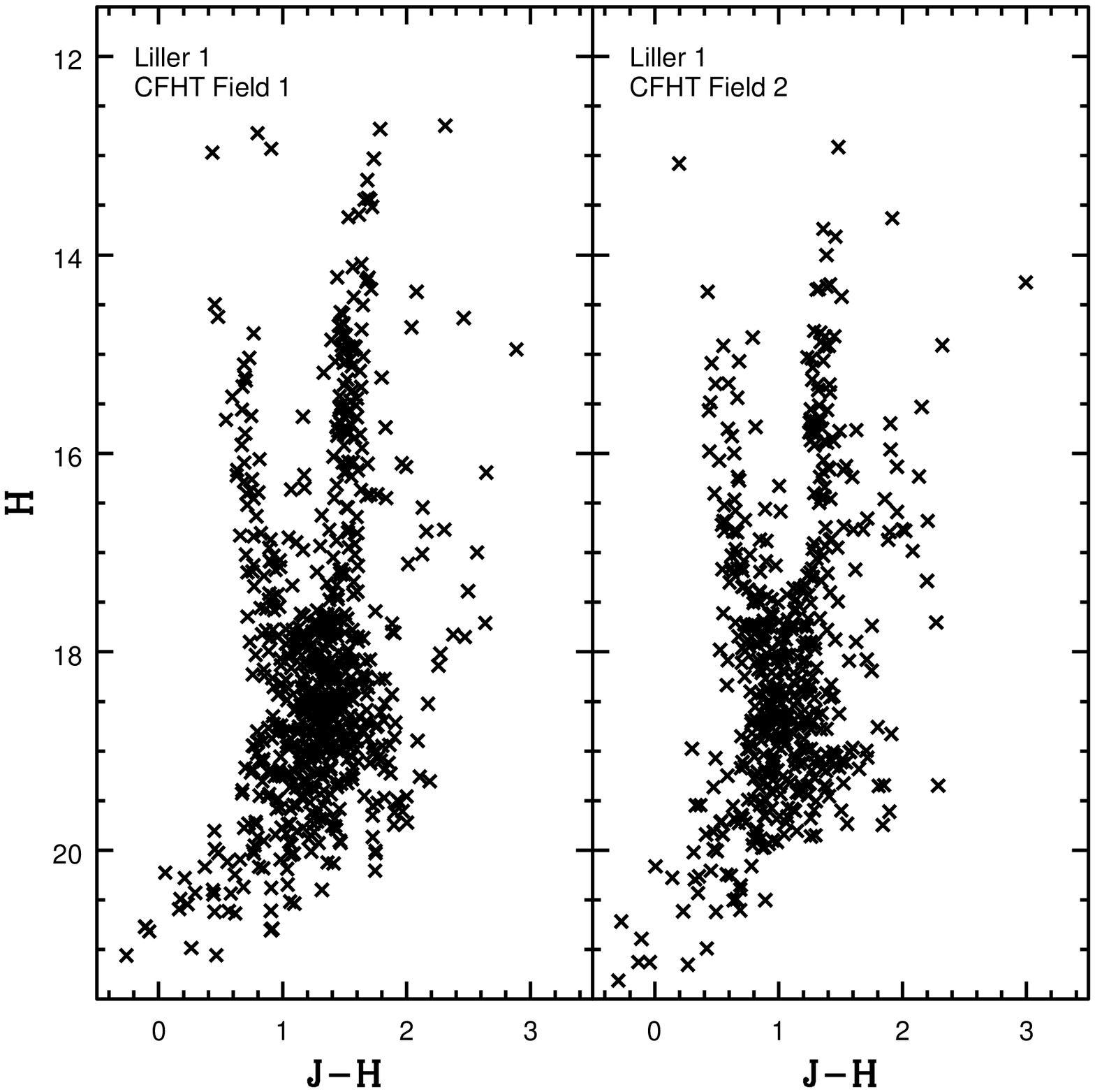]
{The $(H, J-H)$ CMDs of the Liller 1 fields observed with 
AOB+KIR at CFHT. The ratio of cluster to bulge stars on the giant branch is 
roughly 2:1 for Field 1 and 1:1 for Field 2.}

\figcaption
[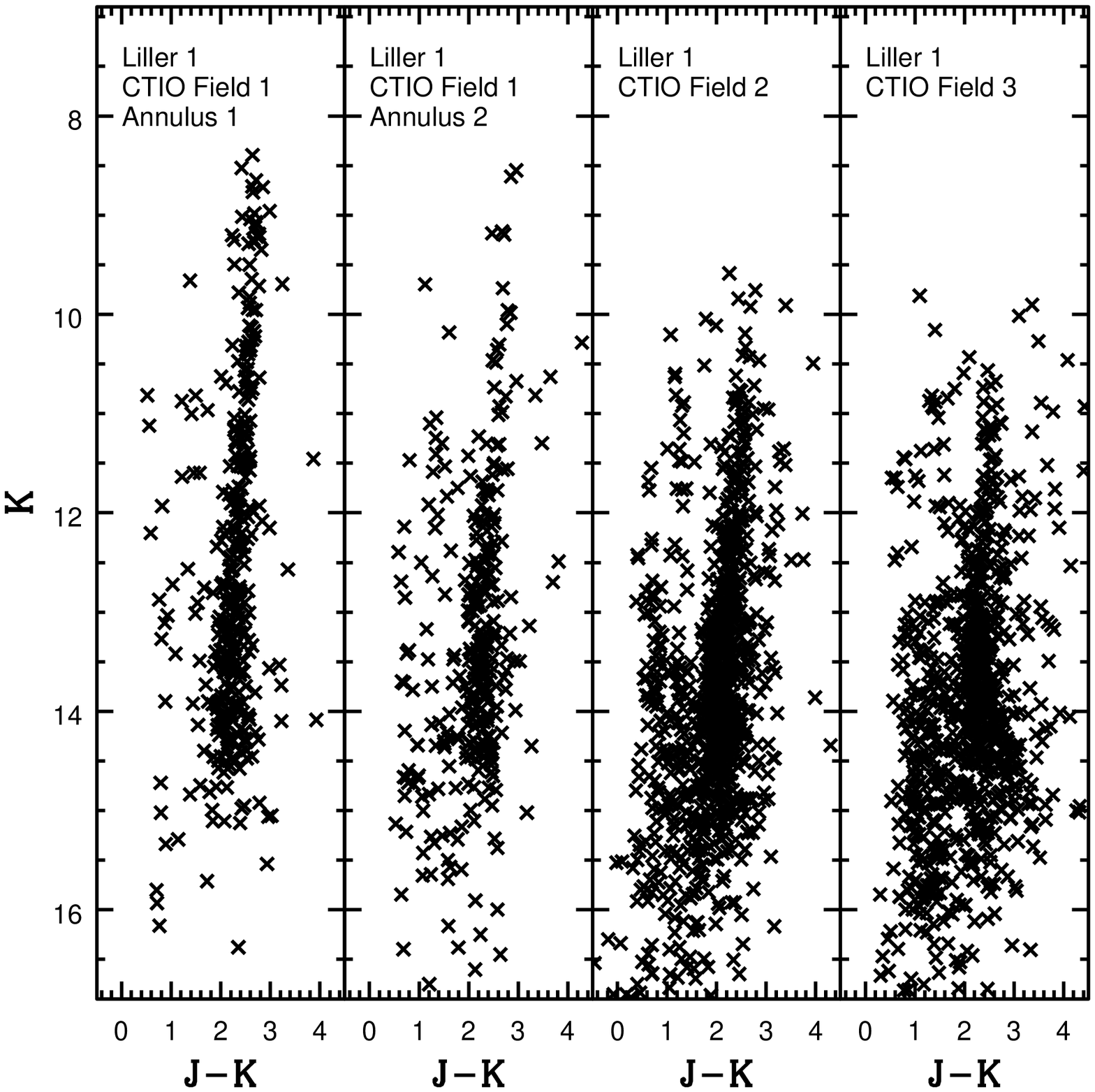]
{The $(K, J-K)$ CMDs of the Liller 1 fields observed with CIRIM at CTIO.}

\figcaption
[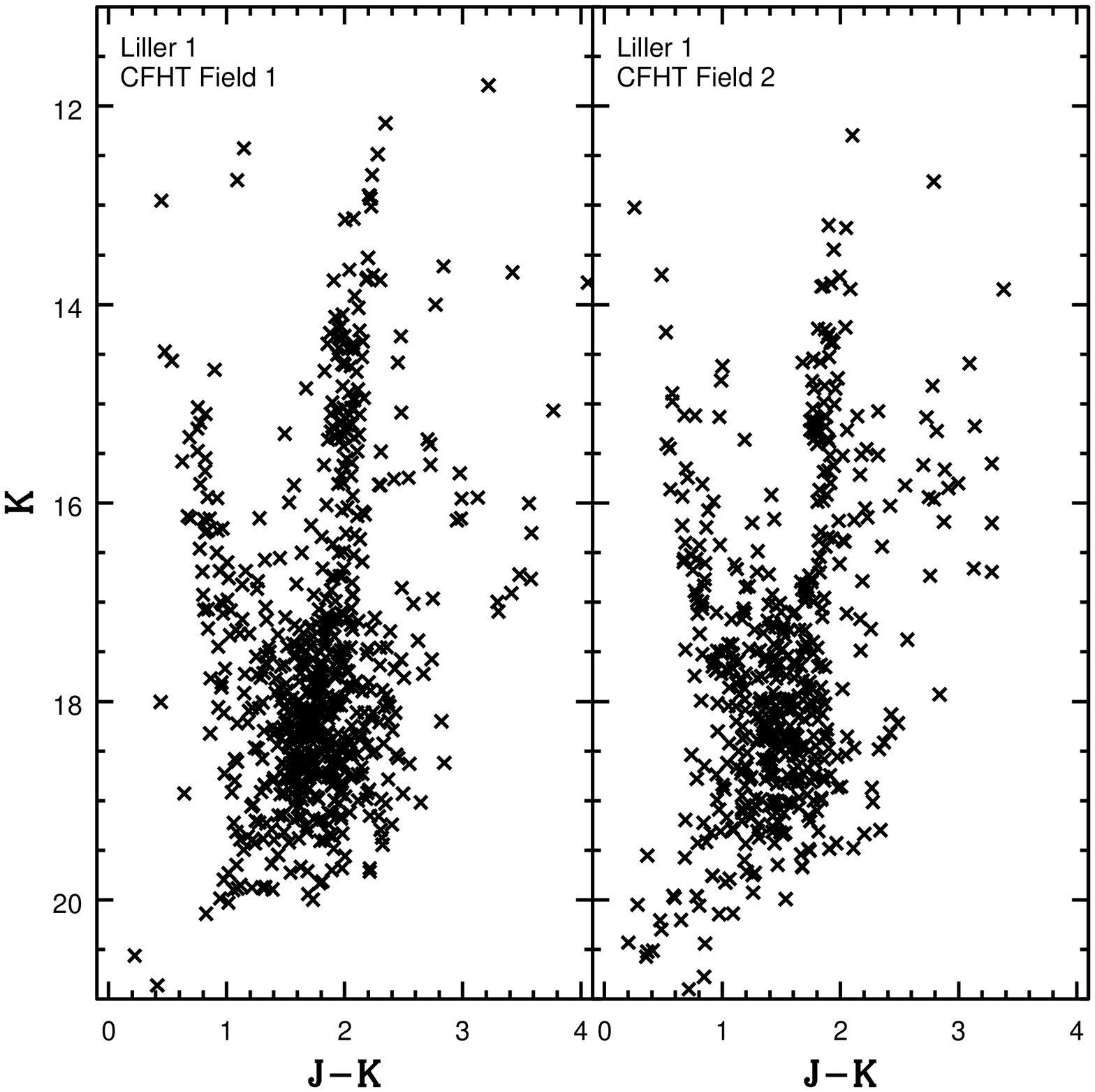]
{The $(K, J-K)$ CMDs of the Liller 1 fields observed with 
AOB+KIR at CFHT. The ratio of cluster to bulge stars on the giant branch is 
roughly 2:1 for Field 1 and 1:1 for Field 2.}

\figcaption
[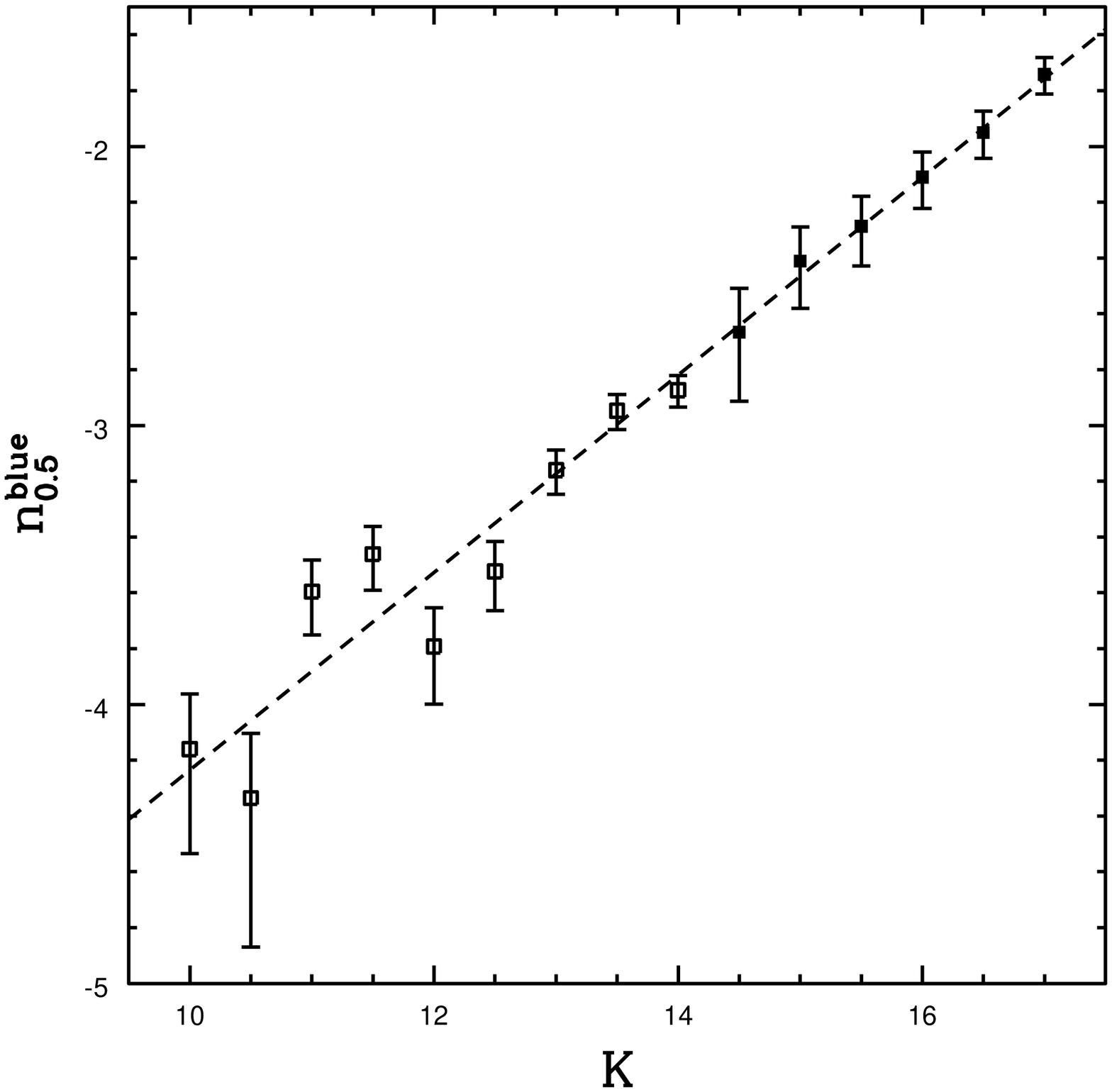]
{The composite $K$ LF of blue plume stars in CTIO Fields 2 and 3 (open 
squares) and CFHT Fields 1 and 2 (filled squares). n$^{blue}_{0.5}$ is the 
number of blue stars per 0.5 mag interval per square arcsec, and 
the dashed line shows a least squares fit to these data. The errorbars show 
the Poisson uncertainties in each bin.}

\figcaption
[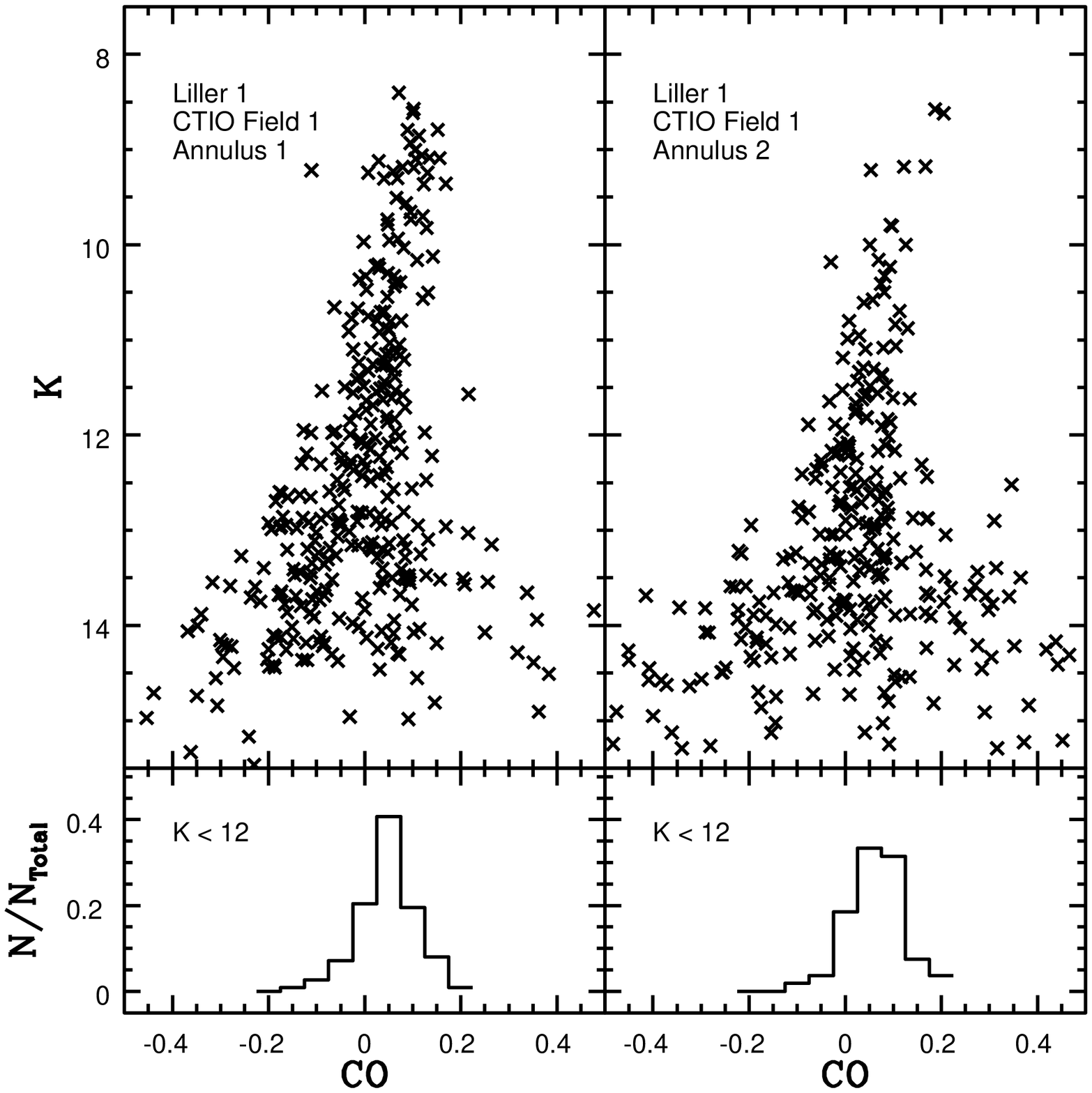]
{The $(K, CO)$ CMDs and histogram distributions of CO 
indices for stars with $K < 12$ in Liller 1 CTIO Field 1 annuli 1 and 2. The CO 
distributions have been normalized to the total number of stars in each 
annulus with $K < 12$. A K-S test indicates that the two CO distributions 
differ at the 97\% significance level.}

\figcaption
[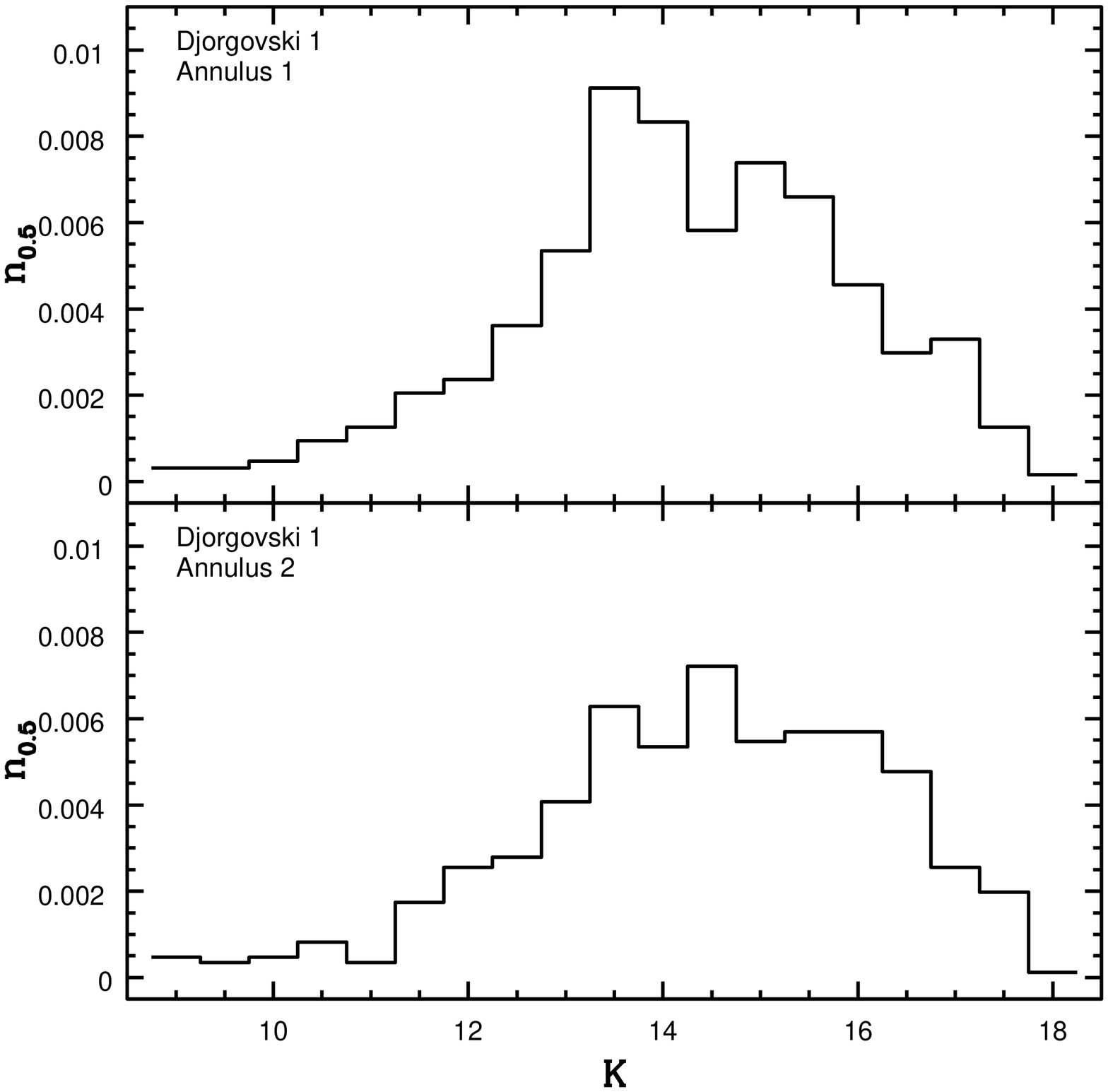]
{The $K$ LFs of Djorgovski 1 annuli 1 and 2, constructed 
from data obtained with CIRIM at CTIO. n$_{0.5}$ is the number of stars per 0.5 
mag interval per square arcsec. Note that the number density of stars in 
annulus 1 only exceeds that of annulus 2 when $K \geq 12.5$, suggesting that 
the most luminous stars in Djorgovski 1 are fainter than the brightest stars 
in the surrounding bulge.}

\figcaption
[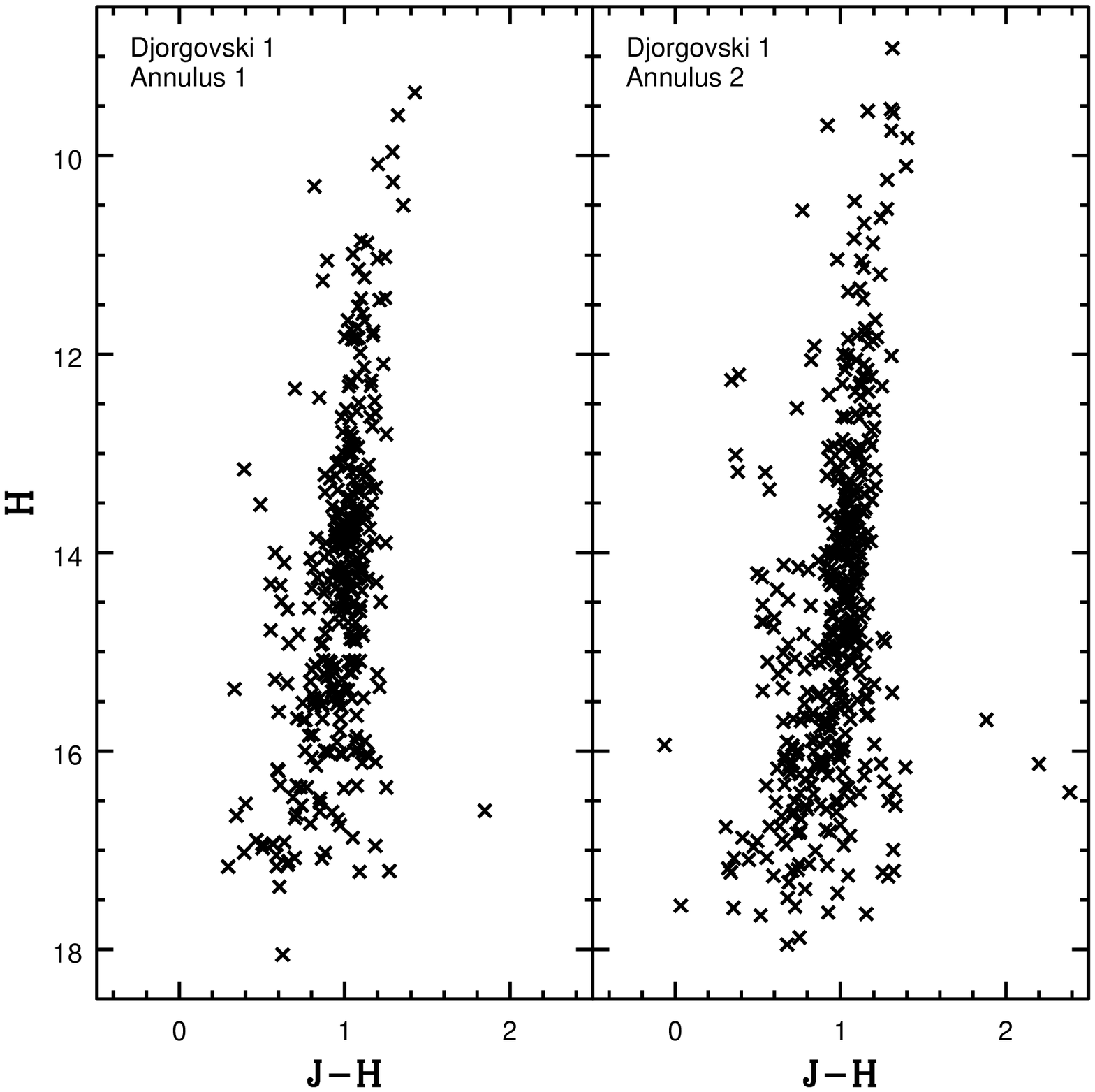]
{The $(H, J-H)$ CMDs of Djorgovski 1 annuli 1 and 2, 
constructed from data obtained with CIRIM at CTIO.}

\figcaption
[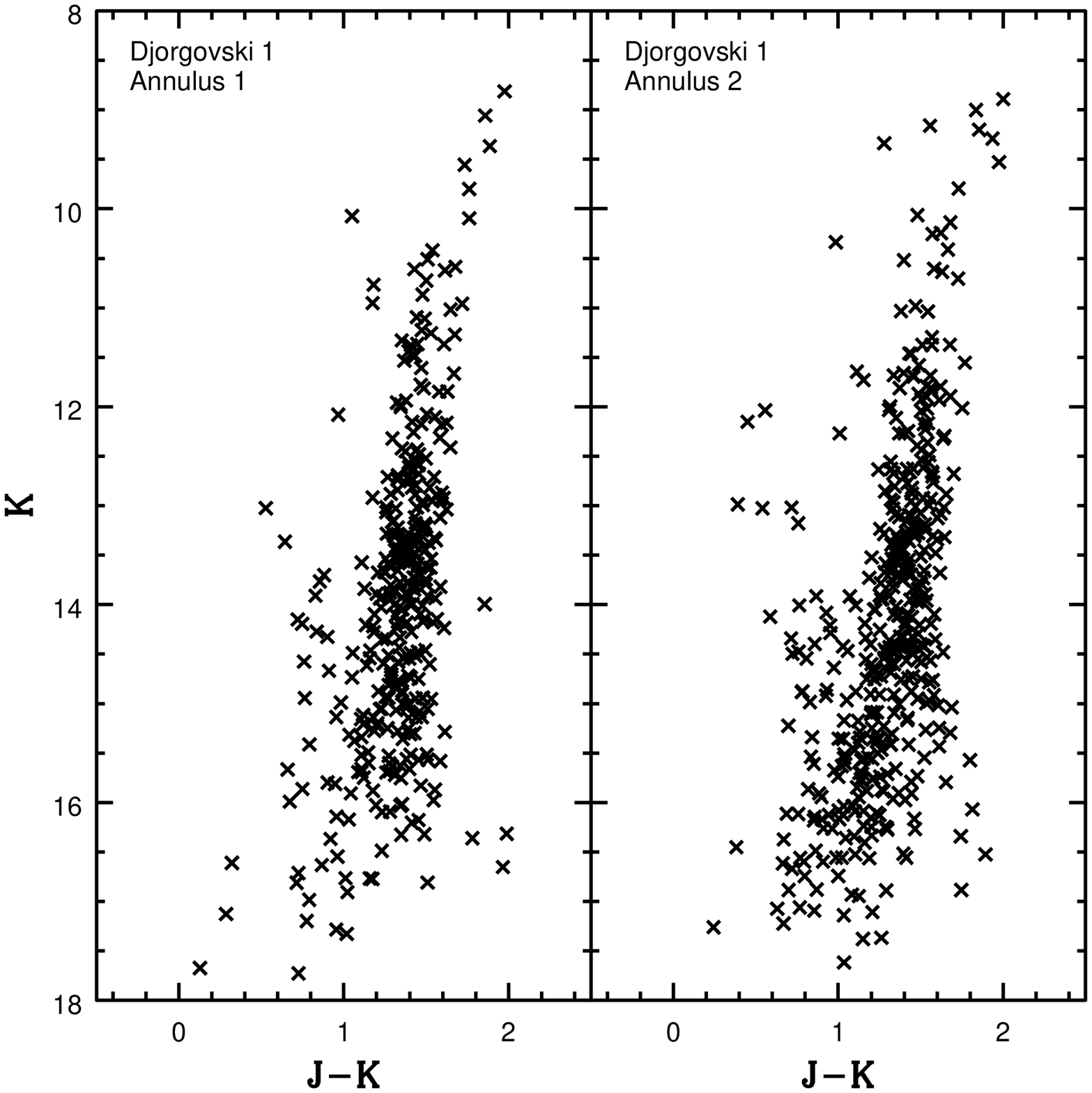]
{The $(K, J-K)$ CMDs of Djorgovski 1 annuli 1 and 2, 
constructed from data obtained with CIRIM at CTIO.}

\figcaption
[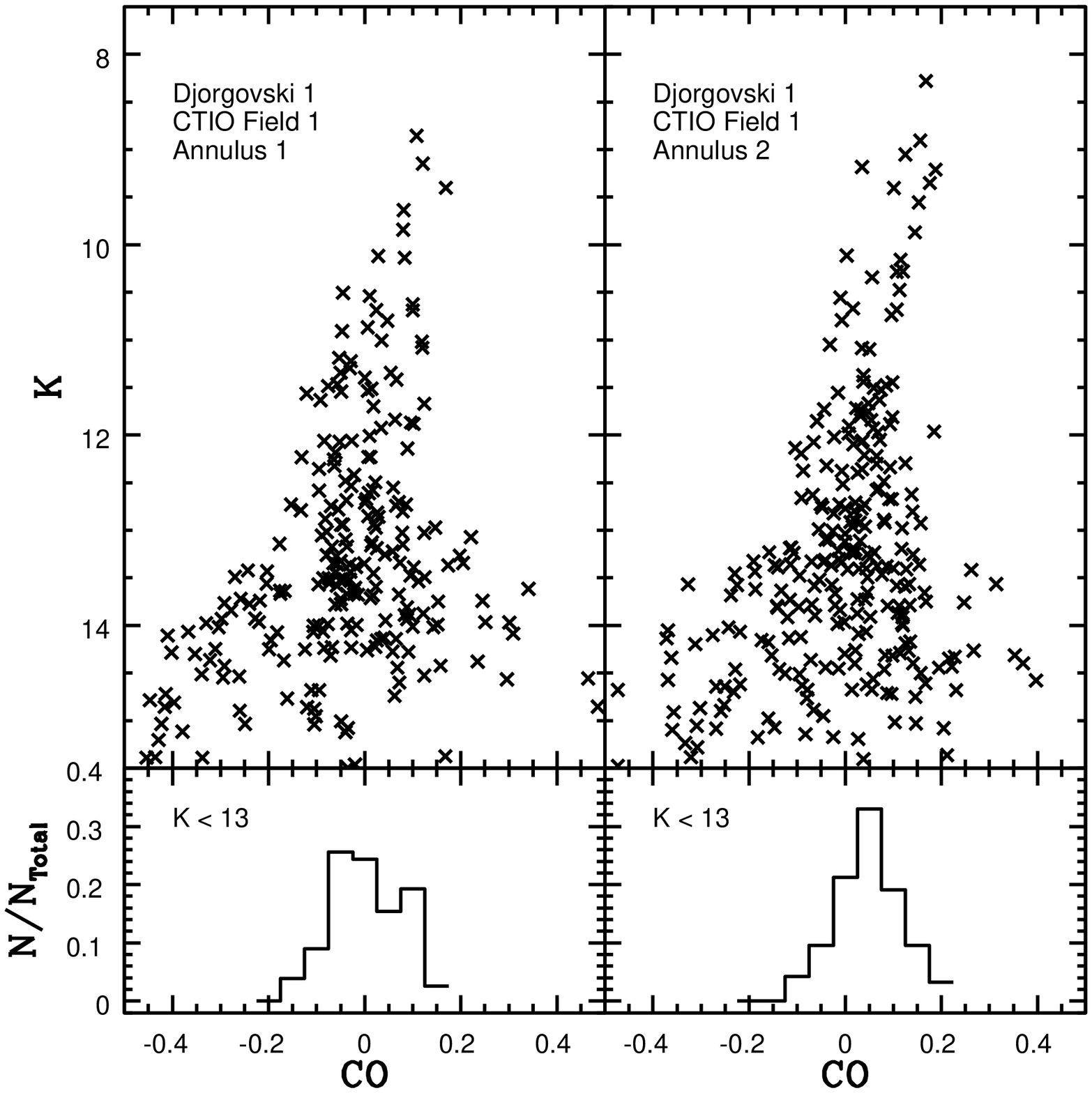]
{The $(K, CO)$ CMDs and histogram distributions of CO indices for stars with 
$K < 13$ in Djorgovski 1 annuli 1 and 2. The CO distributions, which have been 
normalized using the total number of stars in each annulus with $K < 13$, 
differ at the 99\% significance level.}

\figcaption
[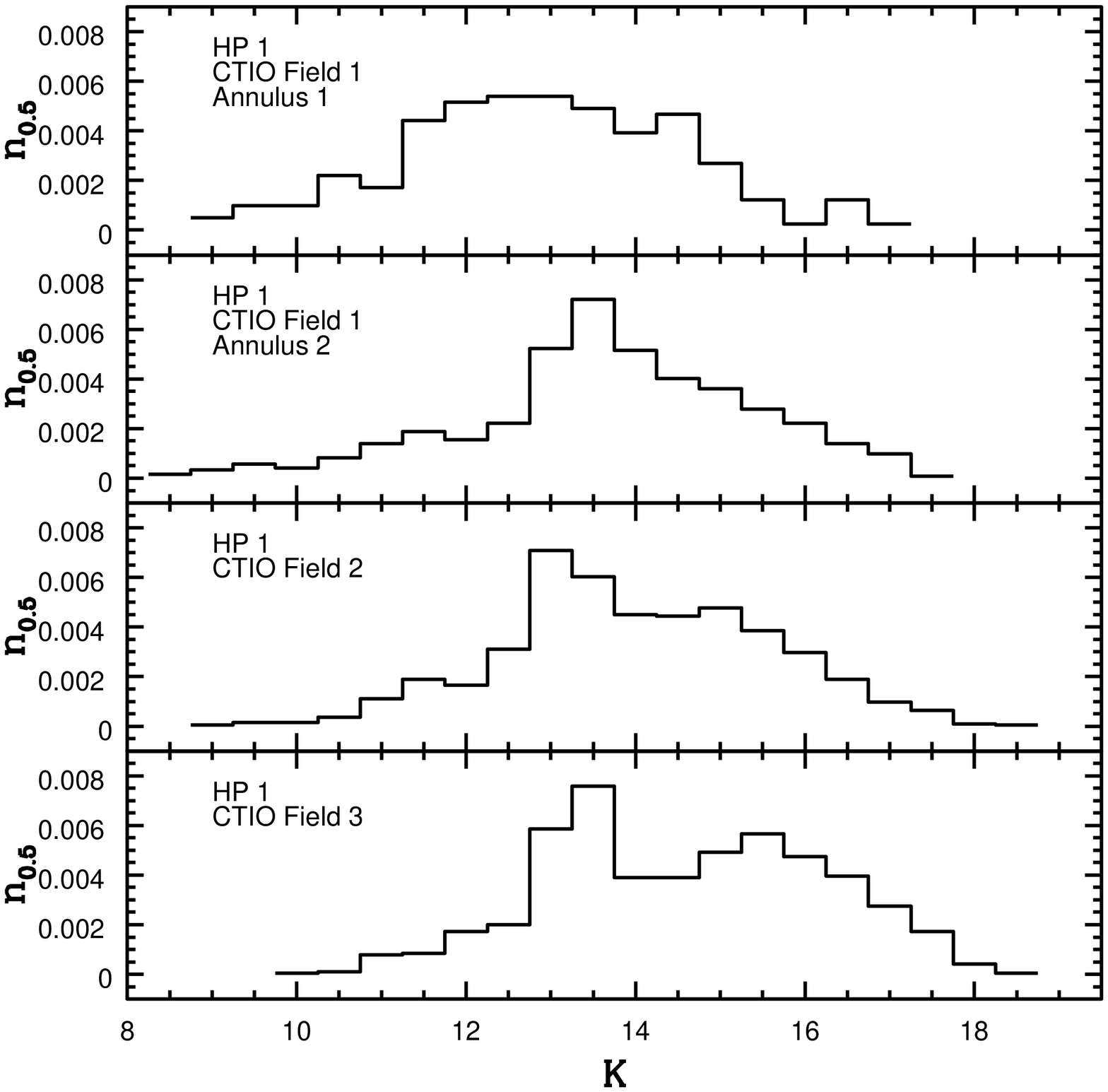]
{The $K$ LFs of HP 1 annuli 1 and 2, constructed 
from data obtained with CIRIM at CTIO. n$_{0.5}$ is the number of stars per 0.5 
mag interval per square arcsec. The number density of stars in 
annulus 1 exceeds that of annulus 2 when $K \geq 10$. The peak in number 
counts when $K = 13 - 13.5$ in Fields 2 and 3 is due to the bulge HB.}

\figcaption
[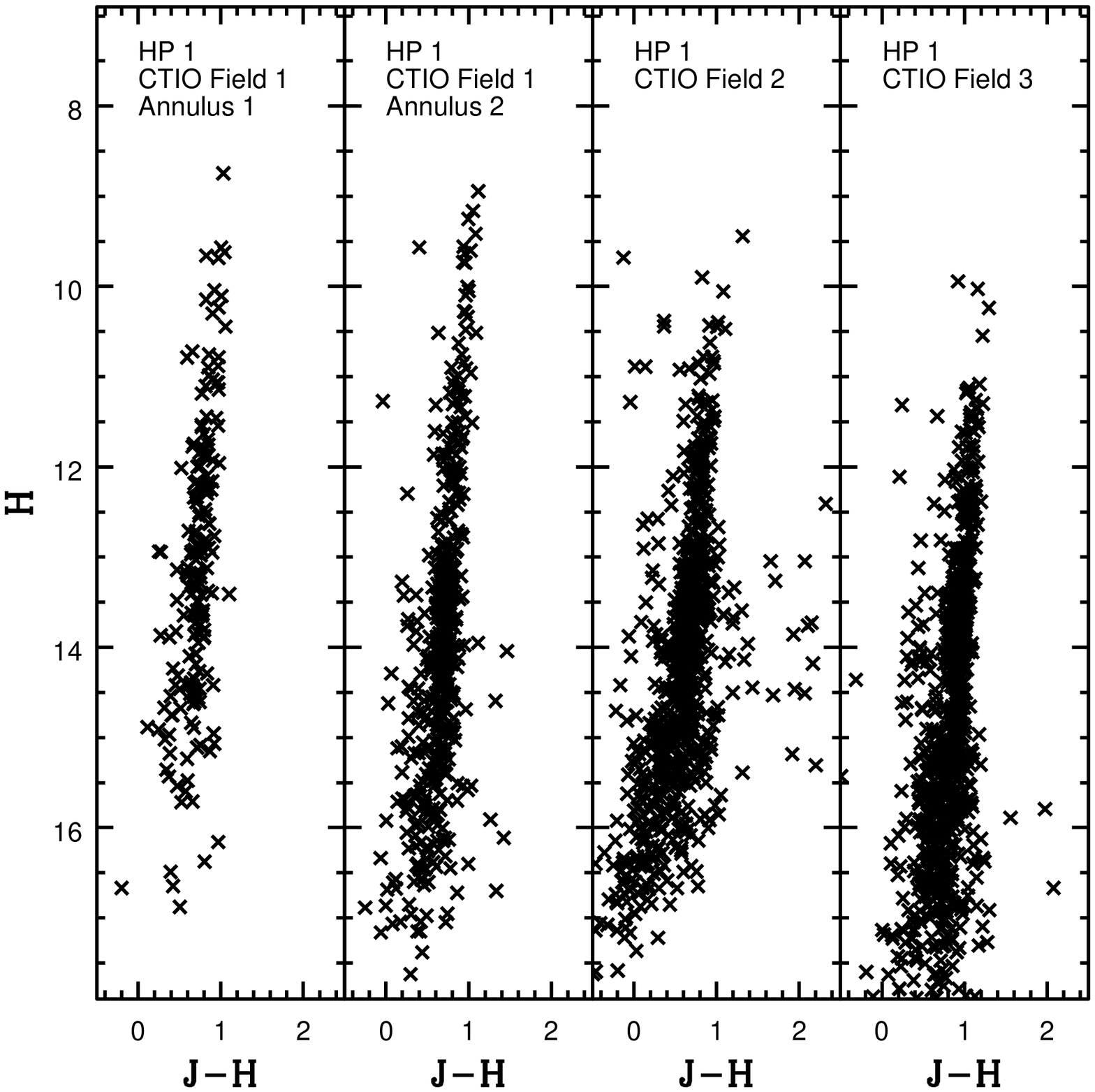]
{The $(H, J-H)$ CMDs of the HP 1 fields, constructed from data obtained with 
CIRIM at CTIO.}

\figcaption
[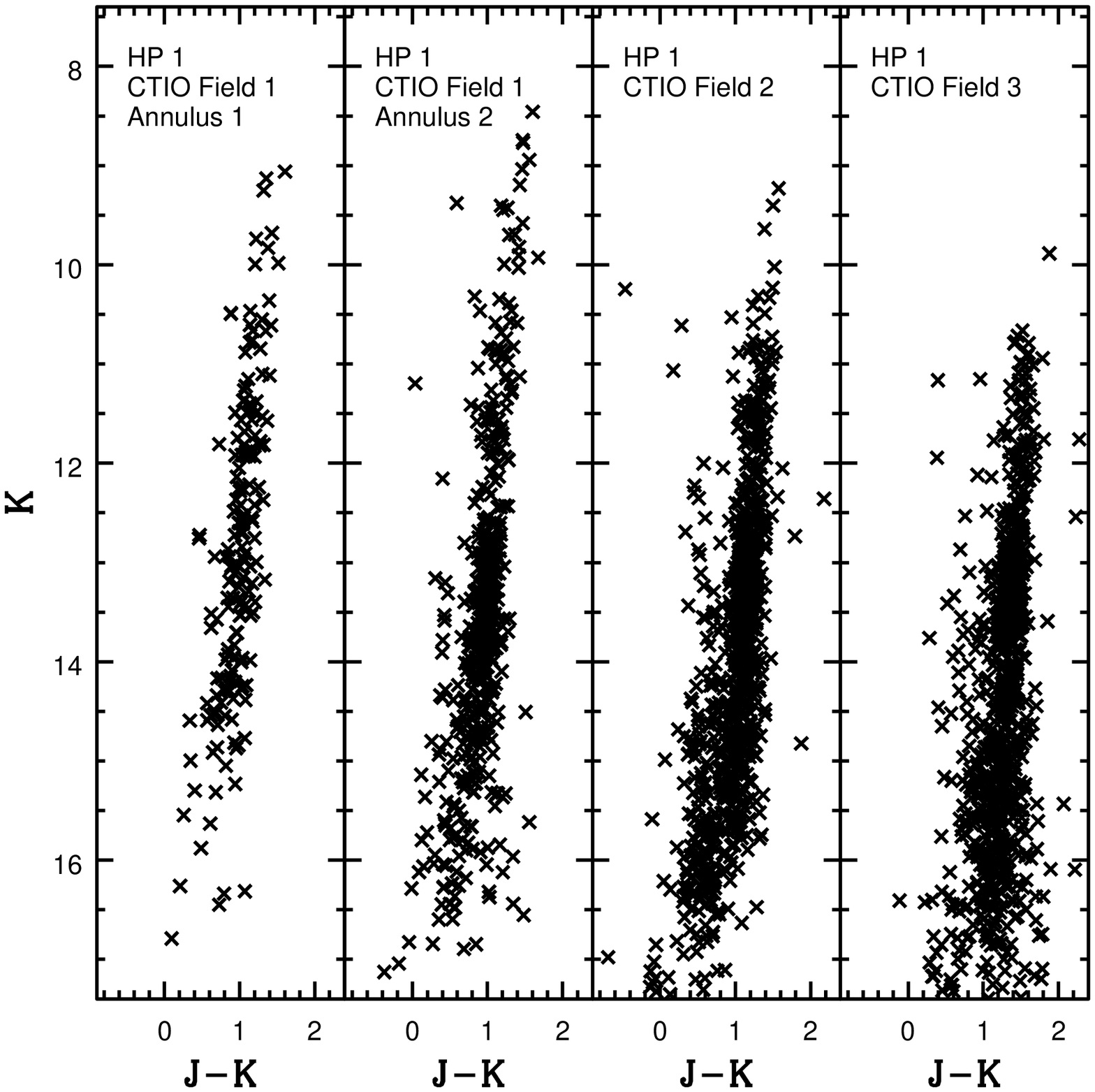]
{The $(K, J-K)$ CMDs of the HP 1 fields, constructed from data obtained with 
CIRIM at CTIO.}

\figcaption
[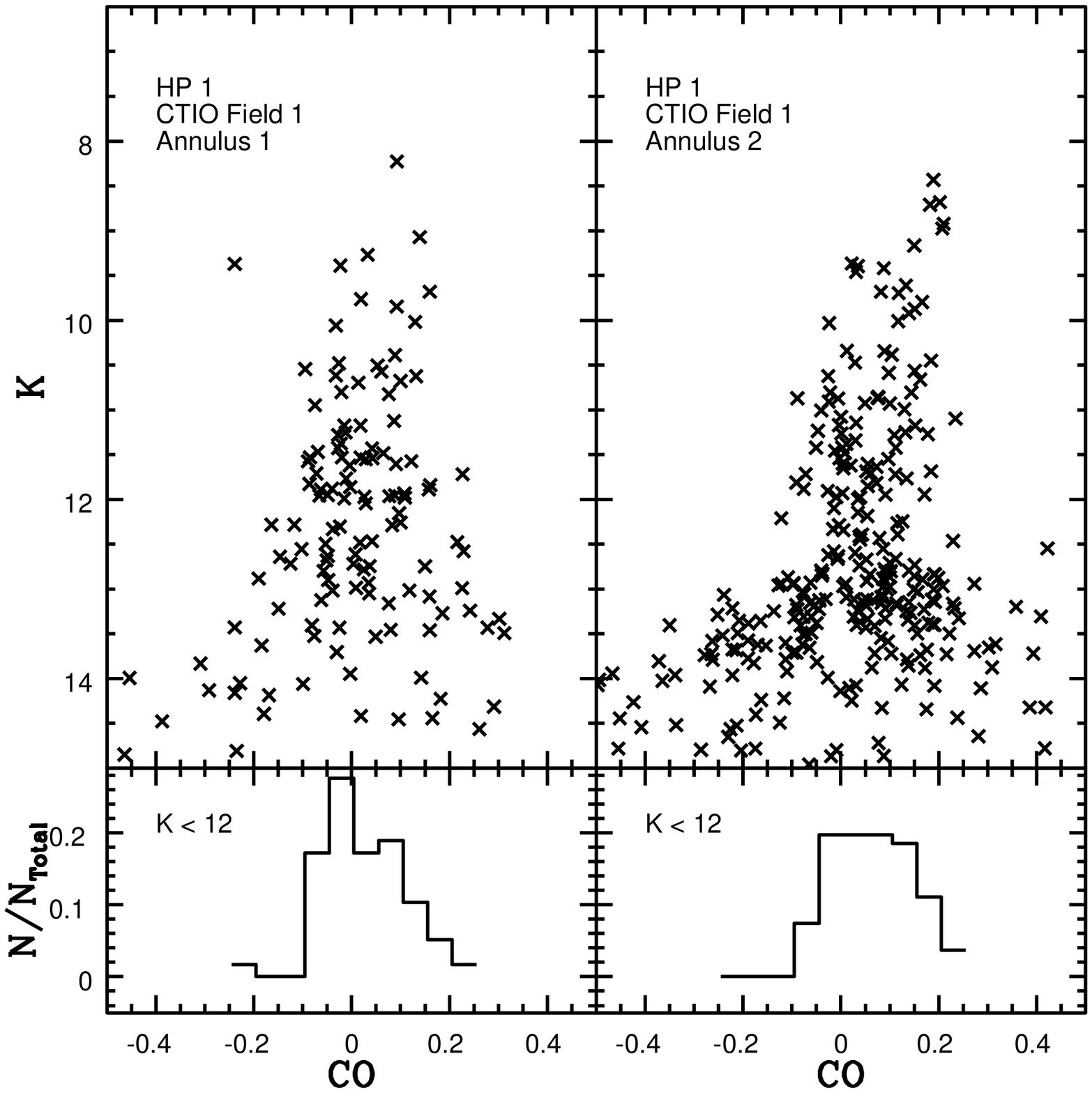]
{The $(K, CO)$ CMDs and histogram distributions of CO indices for stars with 
$K < 12$ in HP 1 annuli 1 and 2. The CO distributions have been normalized 
using the total number of stars in each annulus with $K < 12$. 
A K-S test indicates that the two CO distributions differ at 
the 85\% significance level; however, the means of the distributions differ 
at almost the 99\% significance level.}

\figcaption
[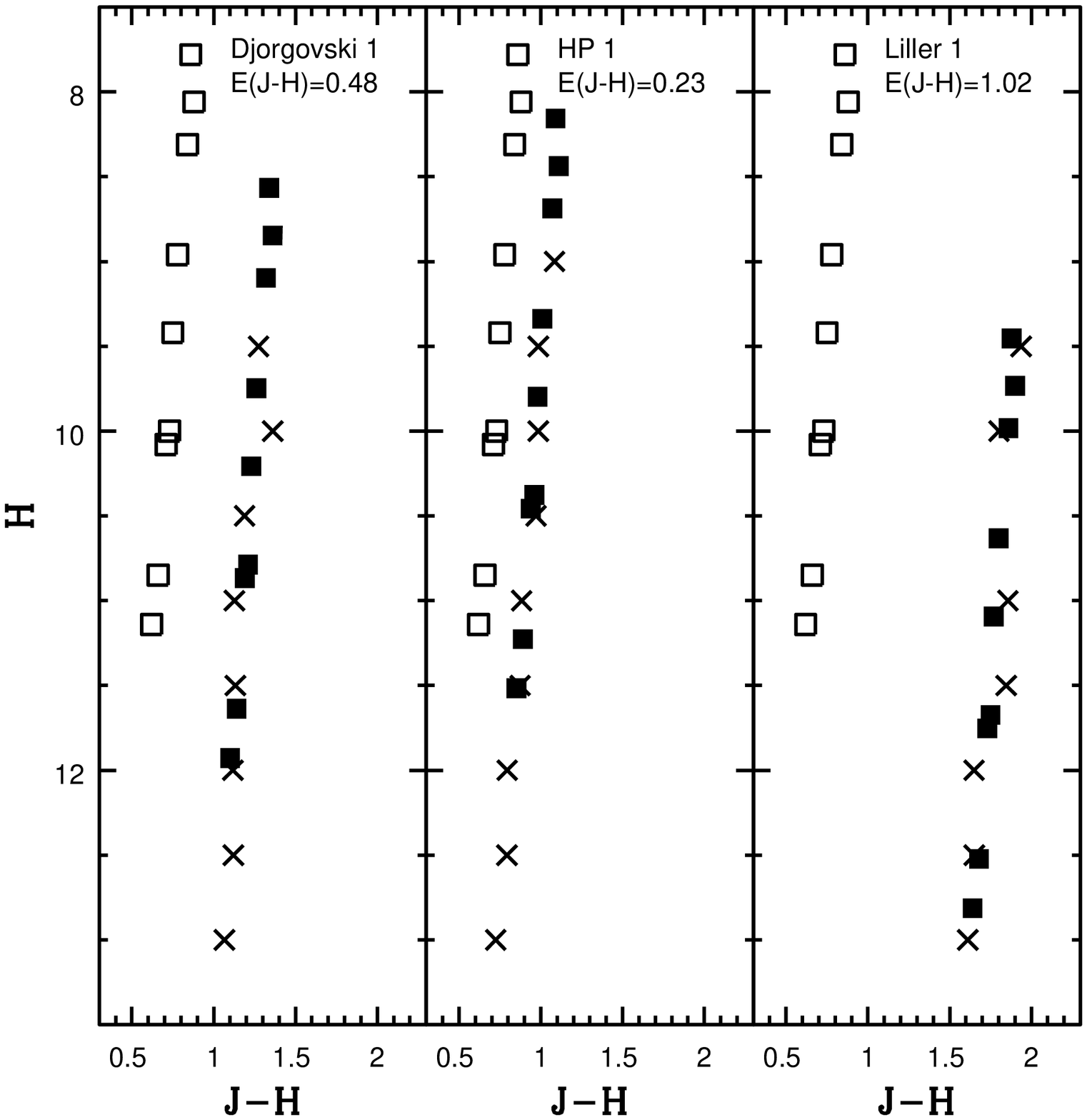]
{Normal points derived from the annulus 2 $(H, J-H)$ 
CMDs (crosses), compared with the BW M giant sequence from Table 3B 
of Frogel \& Whitford (1987). The open squares are the unreddened BW sequence, 
while the filled squares show this sequence shifted along the 
Rieke \& Lebofsky (1985) reddening vector to match the annulus 2 
normal points. The color excess infered for each field is shown at 
the top of each panel.}

\figcaption
[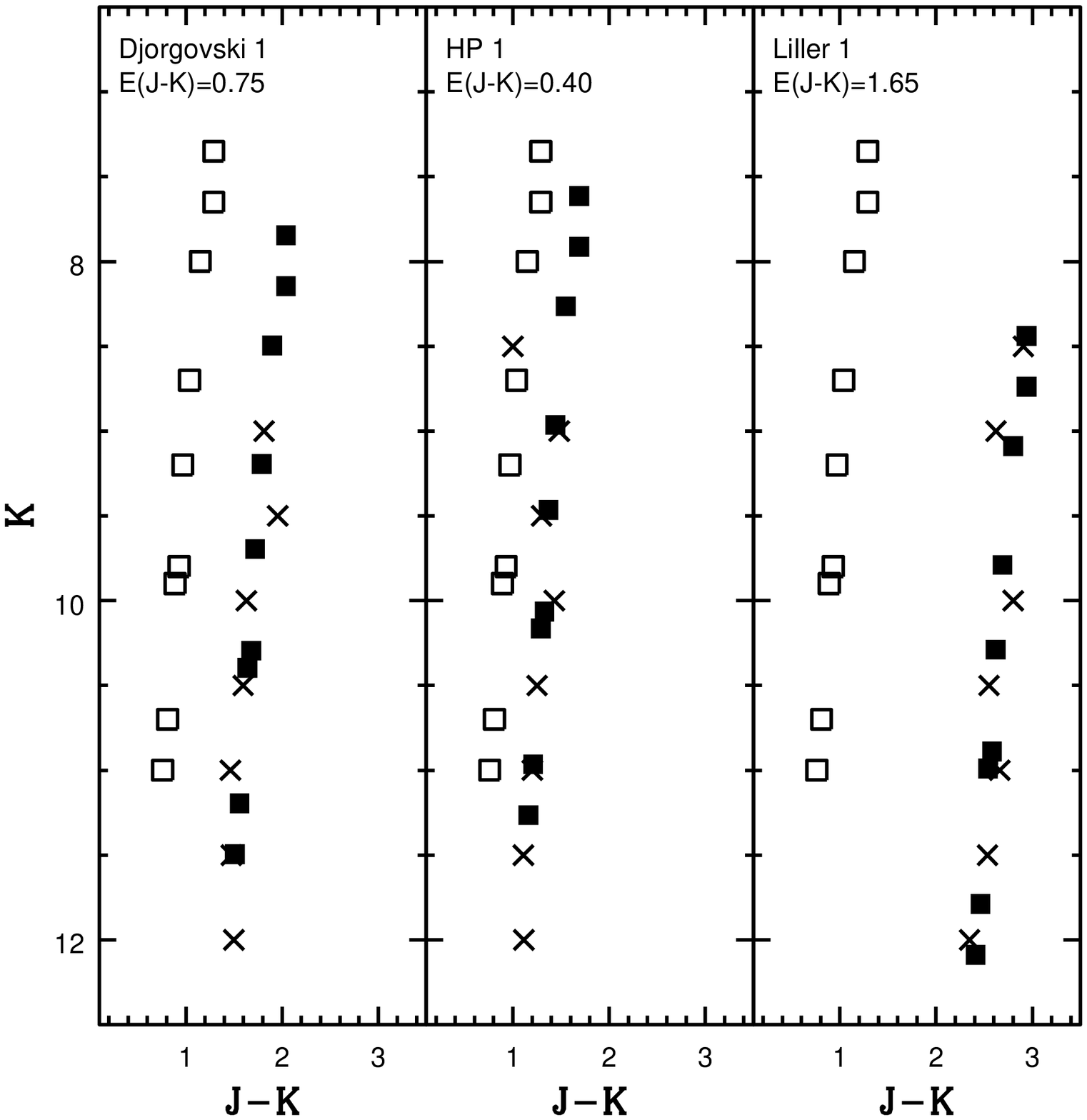]
{Normal points derived from the annulus 2 $(K, J-K)$ 
CMDs (crosses), compared with the BW M giant sequence from Table 3B 
of Frogel \& Whitford (1987). The open squares are the unreddened BW sequence, 
while the filled squares show this sequence shifted along the 
Rieke \& Lebofsky (1985) reddening vector to match the annulus 2 
normal points. The color excess infered for each field is shown at 
the top of each panel.}

\figcaption
[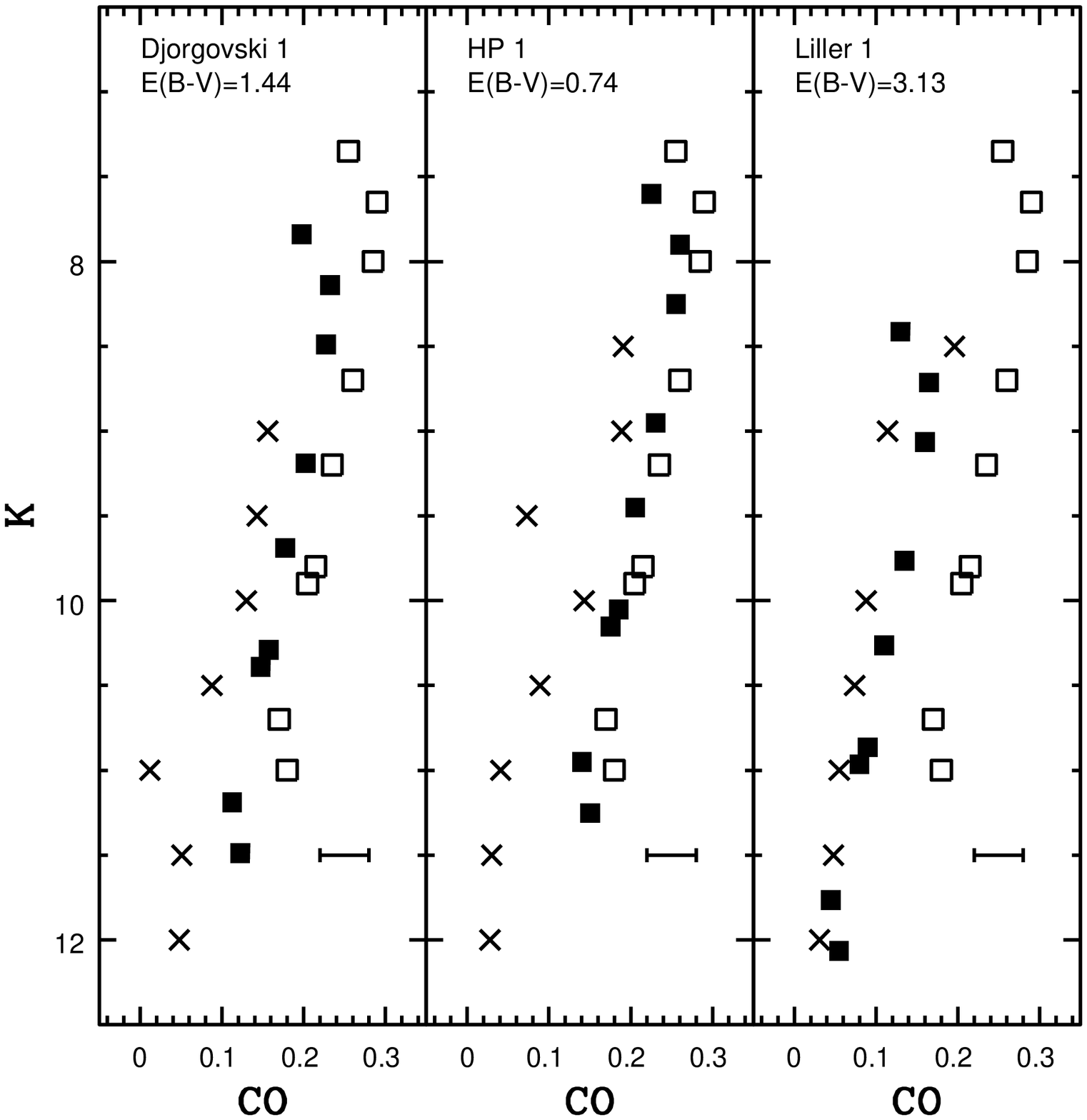]
{Normal points derived from the Annulus 2 $(K, CO)$ 
CMDs (crosses), compared with the BW M giant sequence from Table 3B 
of Frogel \& Whitford (1987). The open squares show the unreddened BW sequence, 
while the filled squares show this sequence shifted along the 
Elias et al. (1985) reddening vector using the $E(B-V)$ values listed 
in the last column of Table 5. The errorbar in the lower right hand corner 
of each panel shows the uncertainty in the photometric zeropoint of the CO 
observations. Note that when $K \leq 10$ the annulus 2 and reddened BW 
sequences for all clusters agree to within the errors in the photometric 
calibration. However, for Djorgovski 1 and HP 1 the agreement is poor when $K 
\geq 10$, likely due to contamination from metal-poor cluster stars.}

\figcaption
[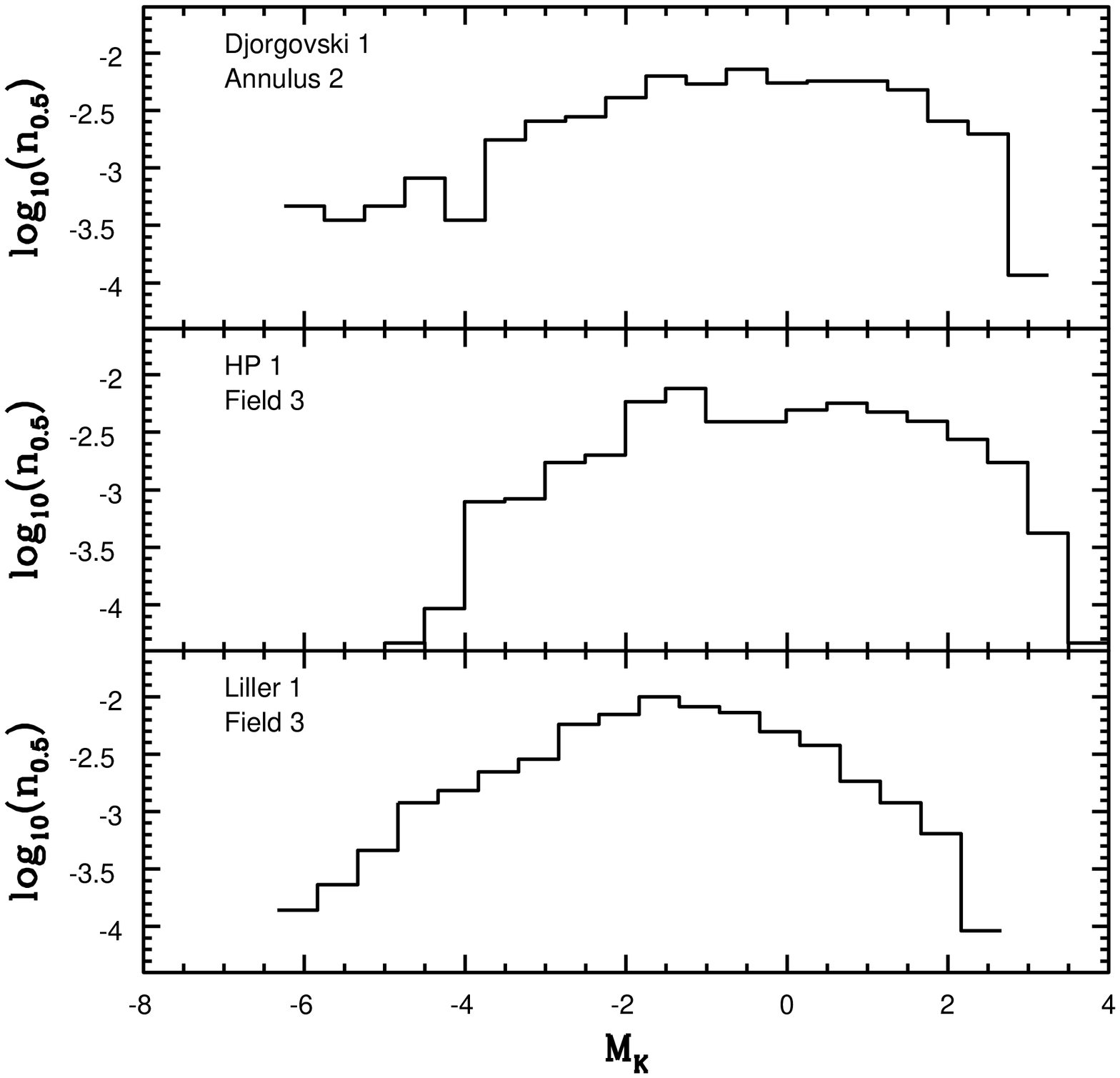]
{M$_K$ LFs for Djorgovski 1 annulus 2, HP 1 Field 3, and Liller 1 Field 3. A 
distance modulus $\mu_0 = 14.5$ (Reid 1993) has been assumed, along with 
the $E(B-V)$ values listed in the last column of Table 5. 
The discontinuity in the HP 1 Field 3 LF at 
M$_K = -4$ defines the bright limit of these data.}

\figcaption
[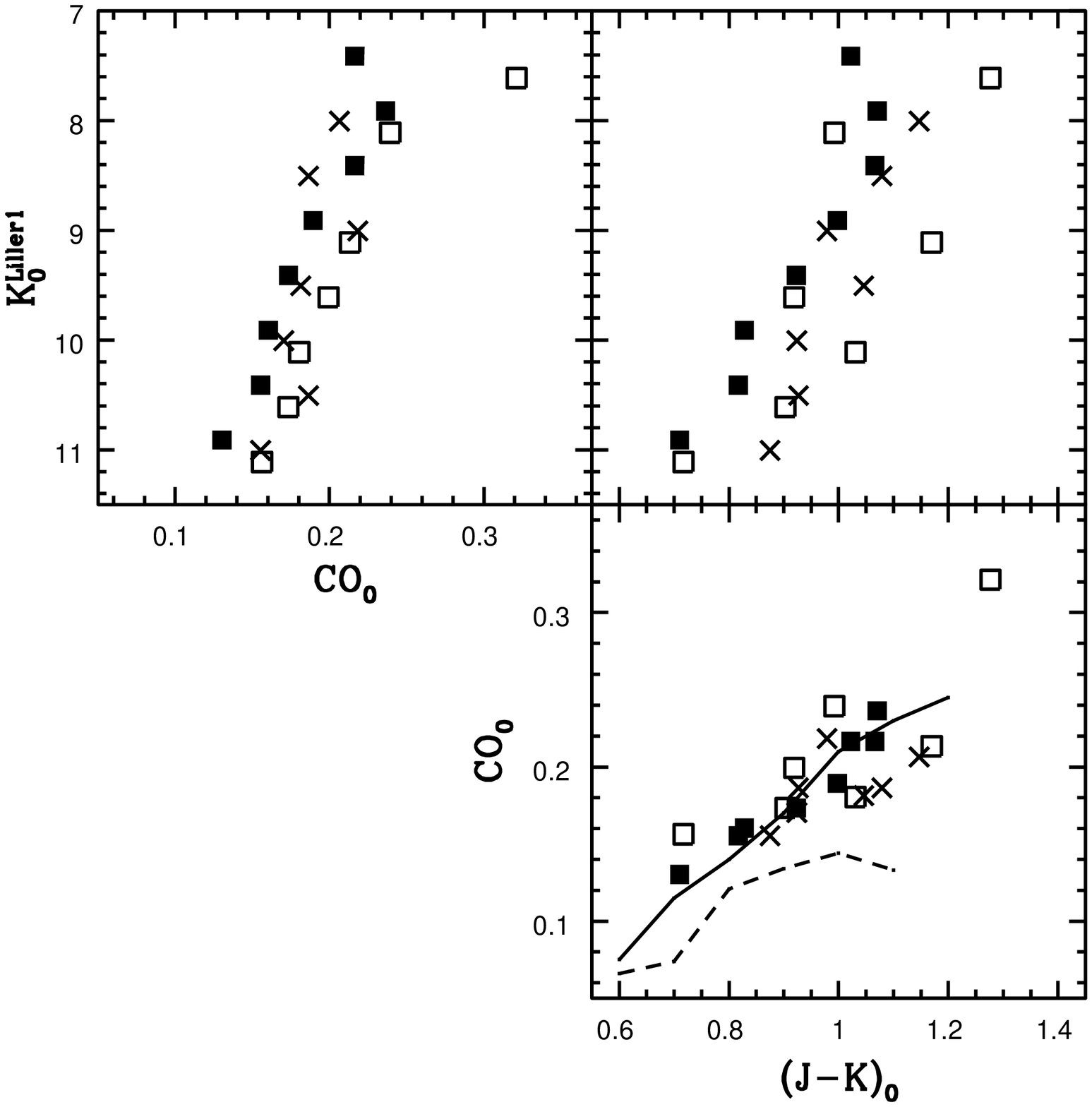]
{(K$_0$, CO$_0$), (K$_0$, (J-K)$_0$), and (CO$_0$, 
(J-K)$_0$) diagrams for Liller 1 (filled squares), the surrounding bulge (open 
squares), and NGC 6528 (crosses). Normal points constructed using the procedure 
described in the text are plotted in this figure. The bulge and NGC 6528 
sequences in the top two panels have been shifted along the $K$ 
axis to match the distance modulus of Liller 1, which is assumed 
to be $\mu_0 = 14.7$ (Frogel et al. 1995). The solid 
line in the lower right hand panel is the field giant sequence from 
Figure 4 of Frogel \& Whitford (1987), while the dashed line shows the trend 
defined by giants in 47 Tuc based on the data tabulated by Frogel et al. 
(1981).}

\figcaption
[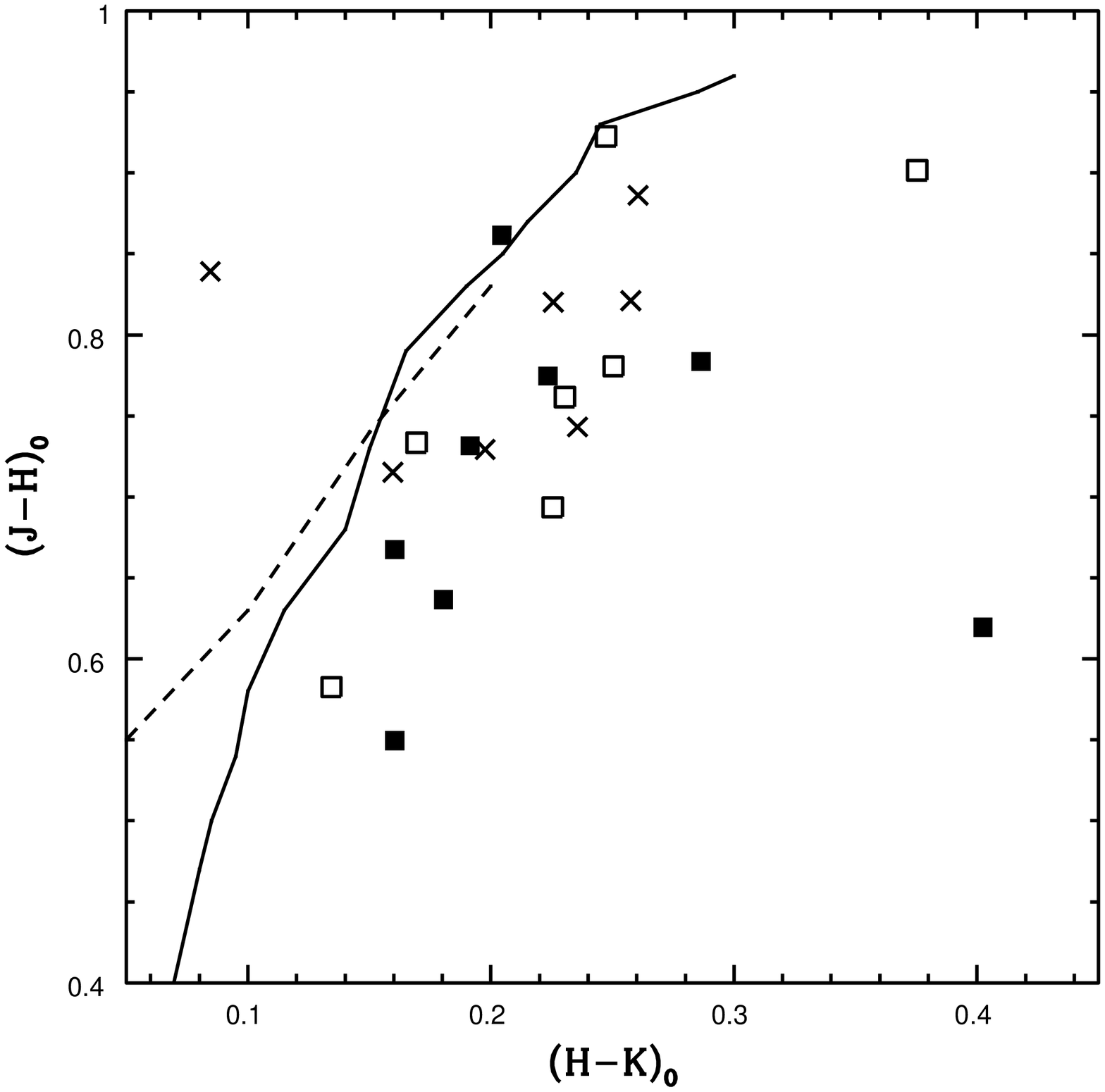]
{The ($J-H, H-K$) TCD for Liller 1 (filled squares), the surrounding bulge 
(open squares), and NGC 6528 (crosses). Normal points constructed using the 
procedure described in the text are plotted in this figure. 
The solid line is the solar neightborhood giant sequence from 
Table III of Bessell \& Brett (1988), while the dashed line shows the 
47 Tuc giant branch sequence, derived from data in Tables 2 and 3 
of Frogel et al. (1981).}

\figcaption
[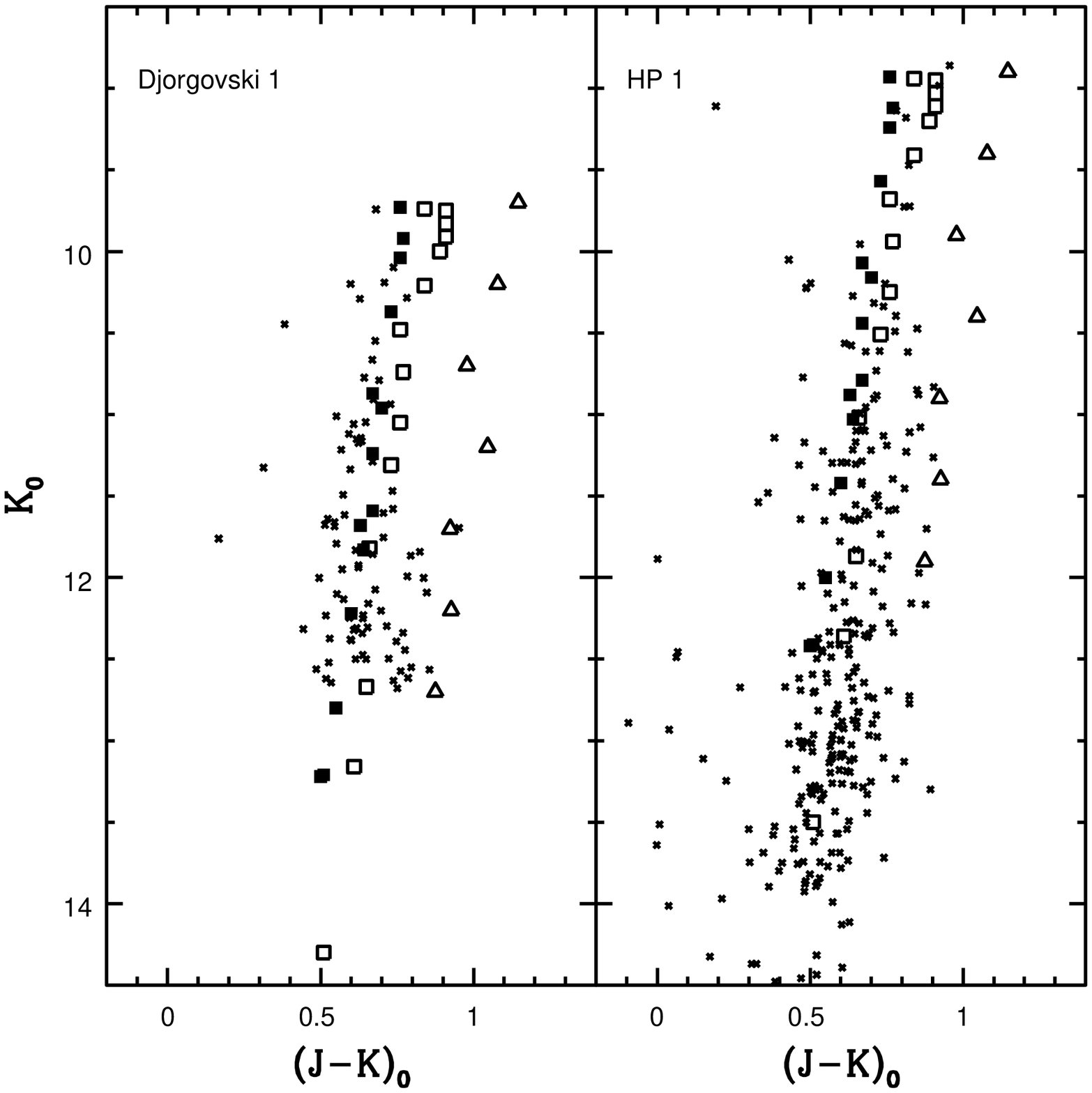]
{The $(K, J-K)$ CMDs of Djorgovski 1 and HP 1 constructed 
from stars selected according to CO index using the procedure described 
in the text. The observations for these clusters have been 
de-reddened. Also shown are sequences for the globular clusters M92 (filled 
squares), M13 (open squares), and NGC 6528 (open triangles), which have 
been de-reddened and registered along the vertical axis with the Djorgovski 1 
and HP 1 data using the brightest stars.}

\figcaption
[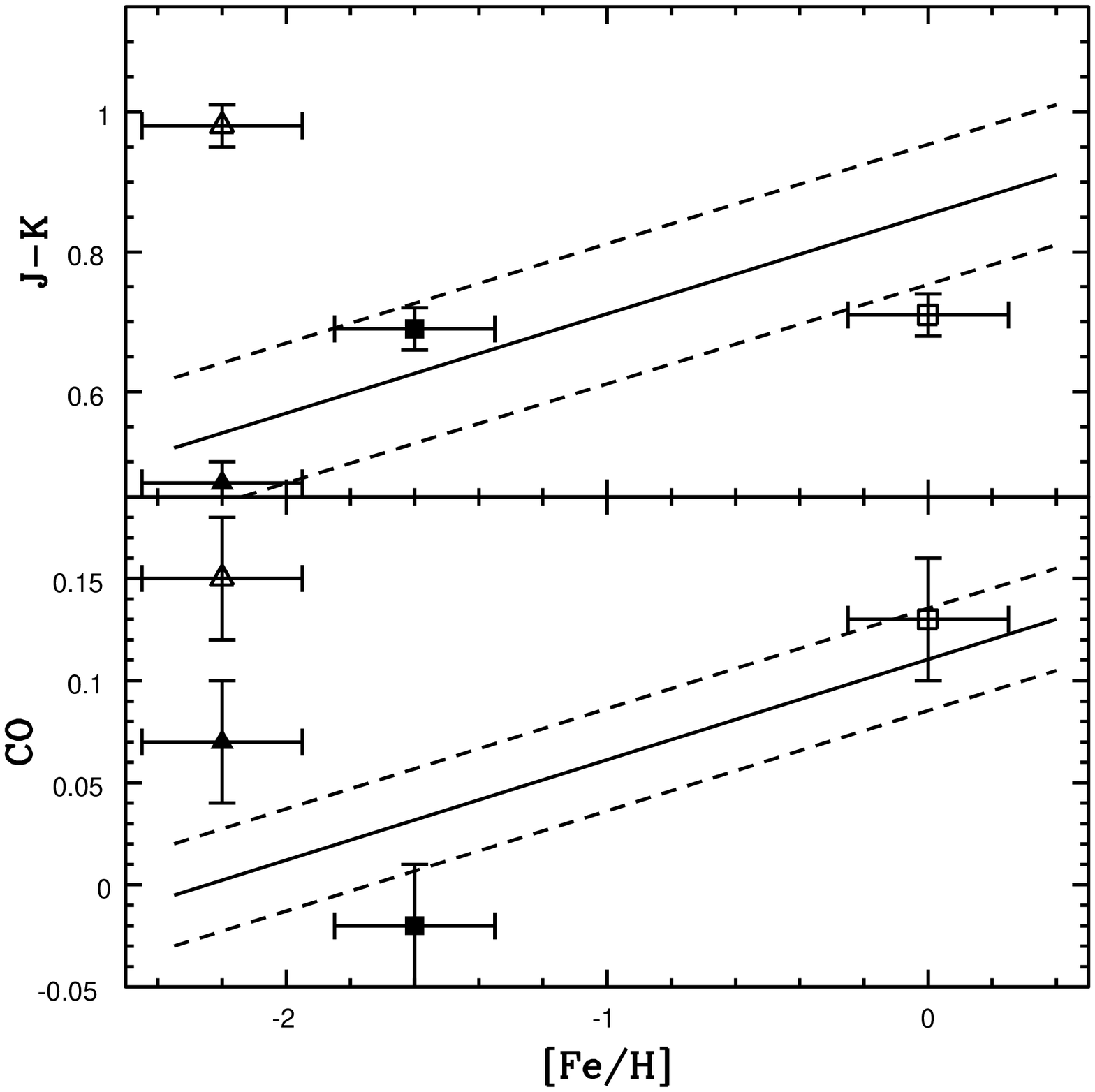]
{The relations between metallicity and integrated color measurements for 
globular clusters. The solid lines are the least squares fits from Figure 1 of 
Aaronson et al. (1978), adjusted to match the Zinn \& West (1984) metallicity 
scale. The dashed lines show the approximate scatter envelop of points plotted 
in Figure 1 of Aaronson et al. (1978). Also shown are data for Liller 1 
(open square) and HP 1 (filled square). The open and solid triangles show the 
measurements for Djorgovski 1 with and without bright field stars. The 
errorbars show the uncertainties in the photometric zeropoints for $J-K$ 
and CO, and assume an estimated $\pm 0.25$ dex uncertainty in the [Fe/H] 
measurements infered from the giant branch ridgelines.}


\clearpage

\begin{table*}
\begin{center}
\begin{tabular}{lrrcc}
\tableline\tableline
Cluster & $l_{II}$ & $b_{II}$ & E(B--V) & R$_{GC}$ \\
 & & & & (kpc) \\
\tableline
Djorg 1 & 356.7 & --2.5 & 1.7 & 1.4 \\
HP 1 & 357.4 & +2.1 & 1.2 & 0.9 \\
Liller 1 & 354.8 & --0.2 & 3.0 & 2.6 \\
NGC 6528 & 1.1 & --4.2 & 0.6 & 0.8 \\
\tableline
\end{tabular}
\caption{CLUSTER PROPERTIES}
\end{center}
\end{table*}


\clearpage

\begin{table*}
\begin{center}
\begin{tabular}{lcccccc}
\tableline\tableline
Cluster & Field \# & RA & Dec & Filter & Exposure Time & FWHM \\
 & & (2000) & (2000) & & (sec) & (arcsec) \\
\tableline
Djorgovski 1 & 1 & 17:47:28.0 & --33:03:59 & J & $4 \times 10$ & 1.2 \\
 & & & & H & $4 \times 10$ & 1.5 \\
 & & & & Ks & $4 \times 10$ & 1.5 \\
 & & & & $2.2\mu$m & $4 \times 20$ & 1.5 \\
 & & & & CO & $4 \times 20$ & 1.5 \\
 & & & & & & \\
HP 1 & 1 & 17:31:05.2 & --29:58:54 & J & $4 \times 10$ & 1.5 \\
 & & & & H & $4 \times 10$ & 1.8 \\
 & & & & Ks & $4 \times 20$ & 1.8 \\
 & & & & $2.2\mu$m & $4 \times 60$ & 1.5 \\
 & & & & CO & $4 \times 90$ & 2.1 \\
 & 2 & 17:31:05.0 & --29:56:56 & J & $4 \times 60$ & 1.5 \\
 & & & & H & $4 \times 60$ & 1.5 \\
 & & & & Ks & $8 \times 60$ & 1.8 \\
 & 3 & 17:31:05.4 & --29:32:57 & J & $4 \times 60$ & 1.5 \\
 & & & & H & $4 \times 60$ & 1.5 \\
 & & & & Ks & $4 \times 60$ & 1.2 \\
 & & & & & & \\
Liller 1 & 1 & 17:33:24.1 & --33:23:23 & J & $4 \times 10$ & 1.5 \\
 & & & & H & $4 \times 10$ & 1.5 \\
 & & & & Ks & $4 \times 20$ & 1.5 \\
 & & & & $2.2\mu$m & $4 \times 10$ & 1.5 \\
 & & & & CO & $4 \times 10$ & 1.5 \\
 & 2 & 17:33:24.2 & --33:21:22 & J & $4 \times 60$ & 1.5 \\
 & & & & H & $4 \times 60$ & 1.5 \\
 & & & & Ks & $4 \times 60$ & 1.5 \\
 & 3 & 17:33:24.2 & --33:08:23 & J & $4 \times 60$ & 1.5 \\
 & & & & H & $4 \times 60$ & 1.5 \\
 & & & & Ks & $4 \times 60$ & 1.5 \\
\tableline
\end{tabular}
\caption{SUMMARY OF CTIO OBSERVATIONS}
\end{center}
\end{table*}


\clearpage

\begin{table*}
\begin{center}
\begin{tabular}{lcccccc}
\tableline\tableline
Cluster & Field \# & RA & Dec & Filter & Exposure Time & FWHM \\
 & & (2000) & (2000) & & (sec) & (arcsec) \\
\tableline
Liller 1 & 1 & 17:33:29.0 & --33:22:53 & J & $4 \times 90$ & 0.24 \\
 & & & & H & $4 \times 90$ & 0.24 \\
 & & & & Ks & $4 \times 90$ & 0.27 \\
 & 2 & 17:33:32.7 & --33:22:36 & J & $4 \times 90$ & 0.31 \\
 & & & & H & $4 \times 90$ & 0.27 \\
 & & & & Ks & $4 \times 90$ & 0.24 \\
 & & & & & & \\
NGC 6528 & 1 & 18:04:48.0 & --30:03:28 & J & $4 \times 5$ & 0.34 \\
 & & & & H & $4 \times 5$ & 0.29 \\
 & & & & Ks & $4 \times 5$ & 0.24 \\
 & & & & $2.2\mu$m & $4 \times 10$ & 0.31 \\
 & & & & CO & $4 \times 10$ & 0.27 \\
\tableline
\end{tabular}
\caption{SUMMARY OF CFHT OBSERVATIONS}
\end{center}
\end{table*}


\clearpage

\begin{table*}
\begin{center}
\begin{tabular}{lccccc}
\tableline\tableline
Cluster & Radius & $J-H$ & $H-K$ & $J-K$ & CO \\
 & (arcsec) & & & & \\
\tableline
Djorgovski 1 & 45 & 1.21 & 0.52 & 1.73 & 0.09 \\
(bright stars & & (0.99) & (0.23) & (1.22) & (0.01) \\
removed) & & & & & \\
 & & & & & \\
HP 1 & 36 & 0.77 & 0.30 & 1.07 & -0.05 \\
Liller 1 & 45 & 1.63 & 0.70 & 2.33 & 0.00 \\
\tableline
\end{tabular}
\caption{APERTURE MEASUREMENTS}
\end{center}
\end{table*}


\clearpage

\begin{table*}
\begin{center}
\begin{tabular}{lccc}
\tableline\tableline
 & & & \\
Cluster & $E(B-V)_{JH}$ & $E(B-V)_{JK}$ & $\overline{E(B-V}$ \\
\tableline
Djorgovski 1 & 1.45 & 1.44 & 1.44 \\
HP 1 & 0.70 & 0.77 & 0.74 \\
Liller 1 & 3.09 & 3.17 & 3.13 \\
\tableline
\end{tabular}
\caption{REDDENING ESTIMATES}
\end{center}
\end{table*}


\clearpage

\begin{table*}
\begin{center}
\begin{tabular}{lcrcr}
\tableline\tableline
Cluster & $(J-H)_0$ & $(H-K)_0$ & $(J-K)_0$ & CO$_0$ \\
\tableline
Djorgovski 1 & 0.73 & 0.25 & 0.98 & 0.15 \\
(bright stars & (0.51) & (--0.04) & (0.47) & (0.07) \\
removed) & & & & \\
 & & & & \\
HP 1 & 0.53 & 0.16 & 0.69 & -0.02 \\
Liller 1 & 0.60 & 0.11 & 0.71 & 0.13 \\
\tableline
\end{tabular}
\caption{DE-REDDENED APERTURE MEASUREMENTS}
\end{center}
\end{table*}

\end{document}